\documentclass[aps,jcp,superscriptaddress]{revtex4-2}

\usepackage{comment}
\usepackage{tikz}
\usetikzlibrary{trees}
\usetikzlibrary{decorations.pathmorphing}
\usetikzlibrary{decorations.markings}
\tikzset{
    photon/.style={decorate, decoration={snake,segment length=1.5mm}, draw=black},
    coulomb/.style={dotted},
    electron/.style={draw=black, postaction={decorate},
        decoration={markings,mark=at position .55 with {\arrow[draw=black]{>}}}}, 
    gluon/.style={decorate, draw=magenta,
        decoration={coil,amplitude=4pt, segment length=5pt}},
    boundelectron/.style={thick, double},
    transverse/.style={dashed}
}

\usepackage{graphicx}
\usepackage{dcolumn}
\usepackage{bm}
\usepackage{rotating}

\newcolumntype{.}{D{.}{.}{8}}

\usepackage[utf8]{inputenc}
\usepackage{physics}
\usepackage{amsmath, amssymb}
\usepackage{tabularx,booktabs,array,dcolumn}
\usepackage{setspace}
\usepackage{vmargin}
\usepackage{xcolor}
\usepackage{graphicx,psfrag,subfigure}

\usepackage{float}
\usepackage[all]{xy}

\usepackage{bm}
\usepackage{mathtools}
\usepackage{physics}
\usepackage[english]{babel}
\usepackage{multirow}

\usepackage{caption}

\newcommand{\bos}[1]{\boldsymbol{#1}}
\newcommand{\mr}[1]{\mathrm{#1}}

\usepackage{tabularx}
\usepackage{dcolumn} 
\newcolumntype{d}[1]{D{.}{.}{#1}}
\usepackage{booktabs}

\def\Eh{E_\mathrm{h}}

\def\eem{\mr{e}}

\def\dd{\mathrm{d}}

\def\br{\bos{r}}
\def\bs{\bos{s}}

\def\be{\bos{e}}
\def\ba{\bos{a}}
\def\bb{\bos{b}}
\def\bc{\bos{c}}
\def\bR{\bos{R}}
\def\bA{\bos{A}}

\def\bE{\bos{E}}
\def\ubE{\bos{\underline{E}}}

\def\bnabla{\bos{\nabla}}
\def\ubA{\bos{\underline{A}}}
\def\bJ{\bos{J}}
\def\ubJ{\bos{\underline{J}}}

\def\som{Supplementary Material}

\usepackage[unicode]{hyperref}
\usepackage{soul}

\def\Eh{E_\text{h}}

\def\eem{\mr{e}}

\def\termS{\text{S}}
\def\Sgp{\Sigma_\text{g}^+}
\def\Sup{\Sigma_\text{u}^+}

\definecolor{ao}{rgb}{0.0, 0.5, 0.0}

\newcolumntype{d}[1]{D{.}{.}{#1}}

\usepackage[unicode]{hyperref}
\hypersetup{
   unicode=true,          
   plainpages=false,
   colorlinks=true,       
   linkcolor=black,          
   linkcolor=blue,          
   citecolor=blue,        
   urlcolor=blue           
}

\usepackage{url}

\begin{document}

\title{%
Regularized relativistic corrections for polyelectronic and polyatomic systems with explicitly correlated Gaussians
}
\author{Bal\'azs R\'acsai} 
\author{D\'avid Ferenc} 
\author{Ádám Margócsy} 
\author{Edit M\'atyus} 
\email{edit.matyus@ttk.elte.hu}
\affiliation{ELTE, Eötvös Loránd University, Institute of Chemistry, 
Pázmány Péter sétány 1/A, Budapest, H-1117, Hungary}

\date{\today}

\begin{abstract}
\noindent %
Drachmann's regularization approach is implemented for floating explicitly correlated Gaussians (fECGs) and molecular systems. 
Earlier applications of drachmannized relativistic corrections for molecular systems were hindered due to the unknown analytic matrix elements of $1/r_{ix}1/r_{jy}$-type operators with fECGs. 
In the present work, one of the $1/r$ factors is approximated by a linear combination of Gaussians, which results in calculable integrals.
The numerical approach is found to be precise and robust over a range of molecular systems and nuclear configurations, and thus, it opens the route towards an automated evaluation of high-precision relativistic corrections over potential energy surfaces of polyatomic systems. 
Furthermore, the newly developed integration approach makes it possible to construct the matrix representation of the square of the electronic Hamiltonian relevant for energy lower-bound as well as time-dependent computations of molecular systems with a flexible and high-precision fECG basis representation. \\[1.5cm]
\end{abstract}

\maketitle

%
\clearpage
The leading-order ($\alpha^2\Eh$) relativistic corrections \cite{BeSabook57} have been computed to high precision for two-, three, and four-electron atoms \cite{Dr81,Dr06,YaDr00,PuKoPa10,PaCeKo05,PuKoPa13,PuPaKo14} and two- and three-electron diatomic molecules \cite{PaCeKo05,CeKoPaSza05,PuKoPa17,JeFeMa21,JeIrFeMa22,FeKoMa20,SaFeMa22}. 
We are aware of a single computation of a polyatomic system (H$_3^+$) for which the regularized leading-order relativistic corrections were computed (at the equilibrium structure) \cite{JeIrFeMa22}.

The terms contributing to the leading-order relativistic correction are large in absolute value and have opposite signs, so the correction values must be converged to several digits to arrive at precise final values. Alternative routes to the mainstream perturbation theory approach have been explored and are promising, potentially also for spectroscopic applications \cite{JeFeMa21,JeFeMa22,FeJeMa22,FeJeMa22b,JeMa23,FeMa23,MaFeJeMa23,MaMa24,NoMaMa24}, but currently available only for two-electron (or two-spin-1/2-fermion \cite{FeMa23}) systems.

Hence, it is relevant to pursue also the traditional approach \cite{BeSabook57} and develop automatization schemes for the precise evaluation of the involved singular operators over generally applicable (explicitly correlated) Gaussian basis sets. Due to the incorrect description of the (electron-electron and electron-nucleus) coalescence points of the non-relativistic wave function represented as a linear combination of explicitly correlated Gaussians (ECGs),
the highly localized, short-range operator expectation values must be regularized to speed up convergence with respect to the basis size \cite{Dr81,PaCeKo05,JeIrFeMa22}. There have been two major directions to correct for this deficiency of general basis sets: 
(a) using operator identities, which replace local operators with global expressions, often called `drachmannization', named after the pioneer of this technique \cite{Dr81};
and (b) integral transformation techniques \cite{PaCeKo05,JeIrFeMa22}, which map the short-range behaviour to the long-range, for which analytic asymptotic expressions can be derived, and can be used to replace the long-tail part of the integral. 

Despite the engaging formal theoretical background~\cite{JeIrFeMa22} of the integral transformation technique~\cite{PaCeKo05}, it is critical to be able to identify a lower threshold value for the long-range part, which is to be represented by the analytic asymptotic expression. Numerical experience shows that a considerable amount of manual fiddling is required to find the optimal threshold value (separating the important physical features represented by numerical integration and the asymptotic limiting function) \cite{FeKoMa20,JeIrFeMa22}. This arbitrariness introduces large uncertainties.

For a precise and automated computation of the leading-order relativistic correction value over a multi-dimensional potential energy surface, potentially at hundreds or thousands of nuclear geometries, a robust methodology is required which can be used in an automated fashion without much manual adjustment.

The drachmannization approach appears to be more robust than the integral transformation technique since it does not contain any parameters to adjust, but the necessary ECG integrals for molecular systems have been thought to be incalculable (in a closed, analytic form) \cite{Ry03}. 
In particular, molecular computations with clamped nuclei can be efficiently carried out using so-called floating explicitly correlated Gaussians (fECGs), in which the shifted (`floating') centres (treated as variational parameters) are necessary to efficiently account for the significant wave function amplitude off-centred from the origin. The $\langle 1/{r_{ix}} 1/{r_{jy}}\rangle$-type integral expressions ($i,j$ label electron, $x,y$ label electron or nuclear coordinates), necessary for the evaluation of the drachmannized relativistic corrections, are not known for fECGs, unless the ECG centres are fixed at the origin (zero shift). 

Interestingly, the same type of incalculable integrals appear in lower-bound theory, and most importantly its Pollak--Martinazzo \cite{MaPo20,PoMa21} version, which has been demonstrated to be powerful for atomic systems (zero-shift ECGs with analytic integrals) \cite{IrJeMaMaRoPo21,RoJeMaPo23}, but no molecular computations have been performed due to the missing integral expressions. 
Potentially, similar integrals will be required in time-dependent simulations recently proposed with explicitly correlated Gaussians \cite{AdKvLaPe22}.

In this work, we develop a numerical approach for the computation of the $\langle 1/{r_{ix}} 1/{r_{jy}}\rangle$-type integrals with fECGs by writing one of the $1/r$ factors as a linear combination of Gaussians. The representation of $1/r$ as a linear combination of Gaussians has already been successfully used for deformed explicitly correlated Gaussians by Varga and co-workers \cite{BeAhHuSuVa21}, using the mathematical scheme of Beylkin and Monz{\'o}n \cite{BeMo05}. 

As a result, a robust numerical drachmannization scheme is developed in this work, and regularized relativistic corrections (computable in a black-box fashion) are reported for two-, three-, and four-electron di- and triatomic molecular systems.

As a starting point, the non-relativistic energy and wave function are computed by solving the
\begin{align}
  H\Psi = E \Psi
  \label{eq:nonrel}
\end{align}
electronic Schrödinger equation with the
\begin{align}
  H 
  = 
  -\sum_{i=1}^n \frac{1}{2} \bnabla^2_{\bos{r}_i} + V
  \label{eq:nonrelH}  
\end{align}
Hamiltonian including the potential energy, 
\begin{align}
  V 
  = 
  \sum_{i=1}^n\sum_{j>i}^n\frac{1}{r_{ij}} 
  -\sum_{i=1}^n\sum_{a=1}^N \frac{Z_a}{|\br_i - \bR_a|} 
  +\sum_{a=1}^N\sum_{b>a}^N \frac{Z_a Z_b}{|\bR_a - \bR_b|} \; ,
\end{align}
where the nuclei have charge number $Z_a$ and are clamped at $\bR_a\ (a=1,\ldots,N$). 
The generally applicable and flexible Ansatz used to represent the wave function is
\begin{align}
  \Psi
  =
  \sum_{\mu=1}^M c_\mu \,
    \mathcal{A} \left\{ \chi_\mu \varphi_\mu \right\} \; ,
    \label{eq:trialfunc1}
\end{align}
where $\mathcal{A}$ is the anti-symmetrization operator (relevant for indistinguishable fermions), and $\chi$ is the spin function written as a linear combination of elementary spin functions coupled to the total electronic spin of the system \cite{PaBook79,MaRe12}.
For the spatial part, we use floating explicitly correlated Gaussian functions (fECGs),
\begin{align}
  \varphi_\mu(\br, \bA_\mu, \bs_\mu) 
  = 
  \exp \left[ -(\br-\bs_\mu)^\text{T} (\bA_\mu \otimes \bos{I}_3) (\br-\bs_\mu) \right] \; ,
\end{align}
where $\br$ collects the $3n$ Cartesian coordinates of the electrons in the system, 
$\bA_\mu \in \mathbb{R}^{n \times n}$ is a positive-definite, symmetric matrix and $\bs_\mu \in\mathbb{R}^{3n}$ will be referred to as the `shift vector'. $\bos{I}_3$ is the three-dimensional unit matrix. The $\bA_\mu$ and $\bs_\mu$ parameters are optimized by minimization of the non-relativistic energy.

In the non-relativistic quantum electrodynamics approach \cite{BeSabook57}, the non-relativistic energy is perturbatively corrected by relativistic and QED contributions in terms of increasing orders of the $\alpha$ fine-structure constant,  
\begin{align}
  E 
  = 
  E^{(0)} + \alpha^2 E^{(2)} + \alpha^3 E^{(3)} + \alpha^3\ln(\alpha){E^{(3,1)}}+ \dots
\end{align}
The $E^{(2)}$ second-order term, known as the leading-order relativistic correction, can be computed as the expectation value of the Breit--Pauli Hamiltonian \cite{BeSabook57}, 
\begin{align}
  E^{(2)} 
  = 
  \langle 
    \Psi 
    | H^{(2)}_{\text{BP}} | 
    \Psi 
  \rangle \; ,
\end{align}
which is written for singlet (or spin averaged) electronic states as
\begin{align}
  H^{(2)}_{\text{BP}} 
  &= 
  -\frac{1}{8} 
  \sum_{i=1}^n 
    \bnabla_i^2 \bnabla_i^2 
  +\frac{\pi}{2}
  \sum_{i=1}^n \sum_{a=1}^N
    Z_a \delta(\br_{ia}) 
  +\pi 
  \sum_{i=1}^n \sum_{j>i}^n 
    \delta(\br_{ij}) + H_\text{OO} 
\label{eq:HBP}
    \\
  &\quad\text{with}\quad
  H_\text{OO}
  =
  -\frac{1}{2} 
  \sum_{i=1}^n \sum_{j>i}^n 
    \frac{1}{r_{ij}} 
    \left(%
      \bos{p}_i \bos{p}_j 
      +
      \frac{\br_{ij} (\br_{ij} \bos{p}_i) \bos{p}_j}{r_{ij}^2} 
    \right) \; .
\end{align}
The terms of $H_\text{BP}^{(2)}$ in Eq.~\eqref{eq:HBP} have traditional names, they are the mass-velocity term, the one-electron Darwin term, the two-electron Darwin term (combined with the spin-spin Fermi contact interaction), and the orbit-orbit term, respectively.
All terms, except the orbit-orbit contribution, include highly localized, singular operators, 
for which the expectation value with the non-relativistic wave function is inefficiently represented over 'smooth' fECG basis functions, \emph{i.e.,} the expectation value converges slowly with respect to the number of basis functions \cite{JeIrFeMa22}.

The drachmannization technique \cite{Dr81} uses operator identities valid for the exact non-relativistic wave function, $(T+V)\Psi=E\Psi$ in Eqs.~\eqref{eq:nonrel}--\eqref{eq:nonrelH}, to replace singular operators with more globally acting operators,
 \begin{align}
  \langle \delta (\br_{ij}) \rangle 
  &=
  \frac{1}{4\pi} 
  \left[%
    2 \langle \Psi | \frac{1}{r_{ij}}(E-V) | \Psi \rangle
    - 
    \sum_{k=1}^n 
     \langle \bnabla_k \Psi | \frac{1}{r_{ij}} | \bnabla_k \Psi \rangle
  \right] \; ,
  \label{eq:Gapprox1}
  \\
  \langle \delta (\br_{ia}) \rangle 
  &=  
  \frac{1}{2\pi} 
  \left[%
    2 \langle \Psi | \frac{1}{r_{ia}}(E-V) | \Psi \rangle
    - 
    \sum_{k=1}^n 
      \langle {\bnabla}_k \Psi | \frac{1}{r_{ia}} | {\bnabla}_k \Psi \rangle
  \right] \; ,
%
\intertext{and}
  \langle \sum_{i=1}^n \bos{\nabla}_i^2 \bos{\nabla}_i^2 \rangle 
  &=
  4\langle \Psi | (E-V)^2 | \Psi \rangle 
  - 
  2\sum_{i=1}^n \sum_{j>i}^n 
    \langle\bnabla_i^2 \Psi | \bnabla_j^2 \Psi \rangle 
  \; .
  \label{eq:Gapprox2}
 \end{align}
During the evaluation of these expectation values, the following integrals appear, 
\begin{align}
  \left\langle \frac{1}{r_{ia}} \frac{1}{r_{kb}} \right\rangle \; , \quad 
  \left\langle \frac{1}{r_{ij}} \frac{1}{r_{kb}} \right\rangle \; , \quad
  \left\langle \frac{1}{r_{ij}} \frac{1}{r_{kl}} \right\rangle \; ,
\end{align}
for which no analytic result is known with fECG basis functions.
In the present work, we propose a numerical approach to compute these integrals to high precision.

\begin{figure} 
  \includegraphics[scale=0.25]{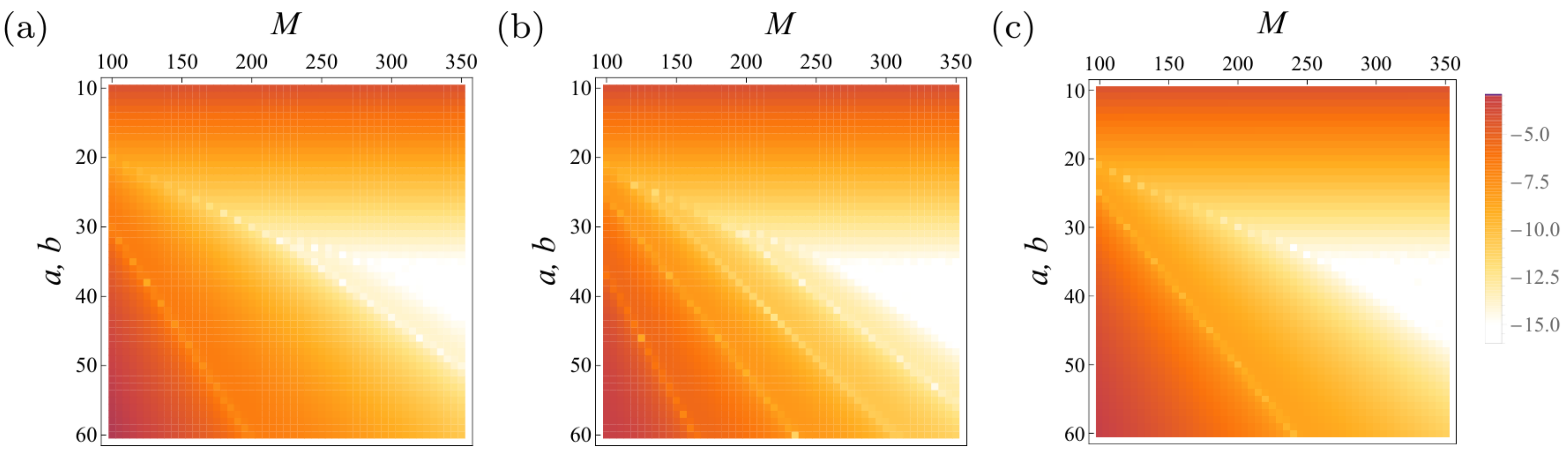}
  \caption{%
    Relative error on a logarithmic scale, $\log |\delta|$, of the Gaussian approximation, Eqs.~\eqref{eq:lingauss}--\eqref{eq:lingaussparam} (with $-a=b$) tested for special cases for which analytic expressions are also available: 
    (a) $U_{kl}=\bra{\varphi_{k}}\frac{1}{r_{ij}}\ket{\varphi_{l}}$;
    (b) $U_{kl}=\bra{\varphi_{k}}\frac{1}{r_{ij}}\frac{1}{r_{pq}}\ket{\varphi_{l}}_{\bs=0}$; 
    (c) $U_{kl}=\bra{\varphi_{k}}\frac{1}{r_{ij}^2}\ket{\varphi_{l}}$.
    $\delta = [U_{kl}^{\text{(analytic)}}-U_{kl}]/U_{kl}$.
    Randomly selected pairs of basis functions were used to prepare the figure (deposited in the \som), and they exemplify the general behaviour observed for all basis sets used in this study. 
    \label{fig:numtest}
  }
\end{figure}

%
We can approximate the $1/r$ function as a linear combination of Gaussian functions. For practical purposes, this approximation is formulated as
\begin{align}
  \frac{1}{r} \cong \sum_{m=1}^M w_m\;\eem^{-p_m r^2} \; ,  
  \label{eq:lingauss}
\end{align}
with the weights and points chosen according to Beylkin and Monzón \cite{BeMo05}, 
\begin{align}
  w_m 
  =
  \frac{2}{\sqrt{\pi}} h
  \eem^{ a + m h } 
  \quad\text{and}\quad
  p_m 
  = 
  \eem^{2 (a+m h)} \; ,
  \label{eq:lingaussparam}
\end{align}
where the step variable is $h=\frac{b-a}{M}$. 
The approximation has three input parameters: 
the $M$ number of Gaussian functions, and the $a<0$ and $b>0$ boundary parameters. 
To better understand the origin of these parameters, we note that the underlying approximation idea starts with considering the integral \cite{HaFaYaBe03,BeMo05}
\begin{align}
    r^{-\alpha}= \frac{2}{\Gamma(\alpha/2)} 
     \int_{-\infty}^\infty \dd s\, \eem^{-r^2\eem^{2s}+\alpha s} \ ,
\end{align}
which can be calculated by the $s=\ln (t/r^2)/2$ substitution.
Then, for $\alpha=1$,
\begin{align}
    \frac{1}{r}= \frac{2}{\sqrt{\pi}} 
     \int_{-\infty}^\infty \dd s\, \eem^{-r^2\eem^{2s}+ s} \ , 
\end{align}
and the integral is approximated using the trapezoidal rule. In this numerical approximation, the lower and upper limits are set to some (finite) $a$ and $b$ values, respectively. The numerical integration limits, $a$ and $b$ are chosen so that the value of the integrand is negligible at the boundaries and beyond them. Then, the application of the trapezoidal rule results in Eqs.~\eqref{eq:lingauss}--\eqref{eq:lingaussparam}. The fast convergence of the trapezoidal rule, motivating this form of the integrand, was pointed out in Ref.~\cite{HaFaYaBe03}.

First, we have tested the Gaussian approximation for three special cases, for which analytic integral expressions are also available. Figure~\ref{fig:numtest} highlights the robustness and excellent performance of the numerical approach. 
As a result of extensive numerical testing, we conclude that high-precision results can be obtained robustly without manual fiddling of the parameters of Eq.~\eqref{eq:lingaussparam}. 
For further computations, we have selected the $M=200$, $a=-31$, and $b=31$ values, which is a computationally well-affordable choice and allows for sufficiently precise results (close to machine precision in double-precision arithmetic) without further adjustment. 
An alternative, similarly useful and practical parameterization was $M=400$, $a=-45$, and $b=45$, for which the computational cost remains low, and was used to test the $(M,a,b)=(200,-31,31)$ parameterization. 
Further refinement of these parameters is possible in the future, but already these choices provided excellent results with a low computational cost and thereby open the route for the computation of relativistic corrections for polyelectronic and polyatomic molecular systems with unprecedented precision in an automated manner.

Tables ~\ref{tab:atomtest} and ~\ref{tab:H2test} present test results of this numerical drachmannization approach for two-, three-, and four-electron atoms and the two-electron hydrogen molecule, for which benchmark-quality data was already available in the literature. 
In these examples, we computed the ground state for each system, but the approach is similarly applicable to excited states, too.

\begin{figure}
  \centering
  \includegraphics[scale=0.5]{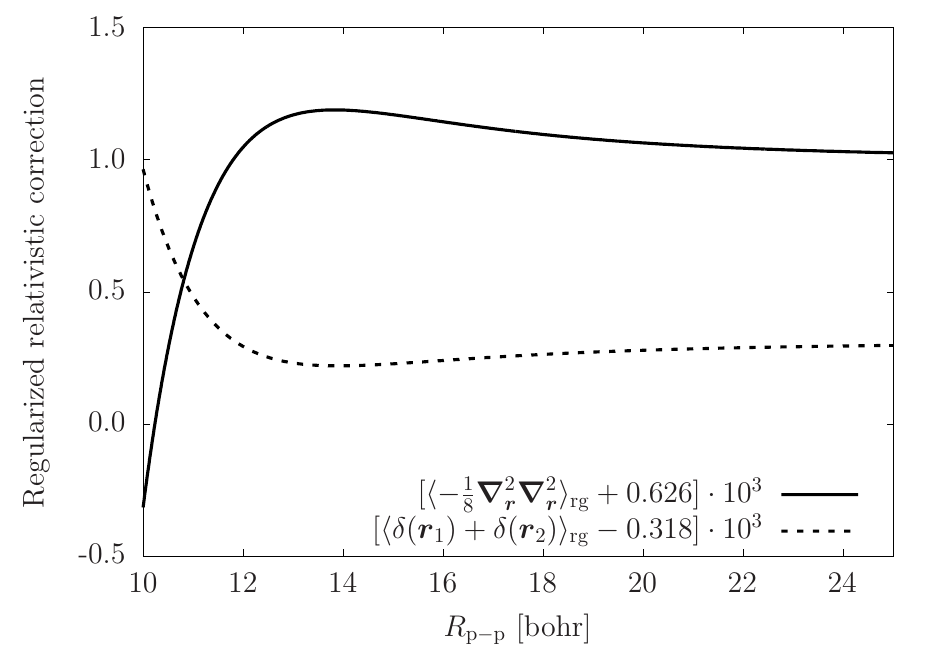}
  \caption{%
    Regularized relativistic corrections to the H$_2^+\ \tilde{A}\ ^2\Sigma_\text{u}^+$ electronic state near the local minimum (near $R_\text{pp}=12.5$~bohr)
    computed by the automated numerical drachmannization approach computed in this work. 
    The computed values confirm (and are more precise) than the ones reported by Kennedy and Howells using special basis functions \cite{HoKe90}.
    \label{fig:h2pa}
  }
\end{figure}

New numerical results for two-, three-, and four-electron, di- and triatomic molecular systems are reported in Table~\ref{tab:molecules}. These results are the first computation of these leading-order relativistic corrections or improve upon the already available values in the literature. 
Furthermore, we report the curve of relativistic corrections to the H$_2^+$ $\tilde{A}\ ^2\Sigma_\text{u}^+$ electronic state (Fig.~\ref{fig:h2pa}) to demonstrate the general applicability of the numerical drachmannization approach along potential energy curves (and potentially, surfaces).

\begin{table}[h]
  \caption{
    Demonstration of the numerical drachmannization approach for the ground state of atoms, 
    He: $1\ ^1\termS$, Li: $2\ ^2\termS$, and Be: $2\ ^1\termS$, 
    and comparison with high-precision reference data from the literature. 
    $\delta E_\text{nr}=E_\text{ref}-E_\text{nr}$ in $\Eh$.
    All computations have been carried out with the same $(M,a,b)=(200,-31,31)$ parameterization of the Gaussian expansion, Eqs.~\eqref{eq:lingauss}--\eqref{eq:lingaussparam}.
    Convergence tables are provided as \som. 
    \label{tab:atomtest} 
  }
\scalebox{0.95}{%
  \begin{tabular}{@{}ll lll@{}}
    \hline\hline\\[-0.40cm]
    \multicolumn{1}{l}{} & 
    \multicolumn{1}{l}{$-\delta E_\text{nr}$} & 
    \multicolumn{1}{l}{$-\frac{1}{8}\sum_i \left\langle \bos{\nabla}_i^2 \bos{\nabla}_i^2 \right\rangle_{\text{rg}}$}  &
    \multicolumn{1}{l}{$\sum_{i,a} Z_a \left\langle \delta(\br_{ia})\right\rangle_\text{rg}$}  &
    \multicolumn{1}{l}{$\sum_{i,j} \left\langle \delta(\br_{ij})\right\rangle_\text{rg}$} 
    \\
    \hline\\[-0.35cm]
    He %
    & $6\cdot 10^{-11}$
    & $-$13.522 054
    & 7.241 717 15
    & 0.106 345 364
    \\    
    He \cite{Dr06} %
    & $1\cdot 10^{-15}$
    & $-$13.522 016 81
    & 7.241 717 274
    & 0.106 345 371  2
    \\        
    \hline \\[-0.35cm]
    Li %
    & $3\cdot 10^{-10}$
    & $-$78.557 5 
    & 41.527 76
    & 0.544 322
    \\
    Li \cite{YaDr00}$^\ast$ %
    & $2\cdot 10^{-13}$
    & $-$78.556 143 062 
    & 41.527 828 927 
    & 0.544 329 79 
    \\    
    \hline \\[-0.35cm]  
    Be %
    & $5\cdot 10^{-7}$
    & $-$270.706 5
    & 141.475 80
    & 1.605 301 5
    \\
    Be \cite{PuKoPa13}$^\ast$ %
    & $3\cdot 10^{-10}$ 
    & $-$270.703 76
    & 141.476 010 4
    & 1.605 305 33
    \\
    \hline\hline
  \end{tabular}
}
  \begin{flushleft}
{\footnotesize
$^\ast$~More precise non-relativistic energy, $E_\text{nr}$, is available from Ref. \cite{PuKoPa10} and \cite{HoAdBu19} for Li and Be, respectively, but we could use the regularized relativistic corrections available from the cited work for comparison. The basis sets for the present computations were taken from earlier work \cite{JeFeMa22,RoJeMaPo23}.
}
  \end{flushleft}
\end{table}

\begin{table}[h]
  \caption{
    Illustration of the enhanced convergence rate by the numerical drachmannization approach developed in this work for the example of the H$_2$ molecule 
    ($\text{X}\ ^1\Sgp$, $R_\text{p-p}=1.40$~bohr).
    $\delta E_\text{nr}=E_\text{ref}-E_\text{nr}$ in $\Eh$.
    All computations have been carried out with the same $(M,a,b)=(200,-31,31)$ parameterization of the Gaussian expansion, Eqs.~\eqref{eq:lingauss}--\eqref{eq:lingaussparam}.
    \label{tab:H2test} 
  }
  \begin{tabular}{@{}ll lll@{}}
    \hline\hline\\[-0.35cm]
    $-\delta E_\text{nr}$ &
    \multicolumn{1}{l}{$-\frac{1}{8}\sum_i \left\langle \bos{\nabla}_i^2 \bos{\nabla}_i^2 \right\rangle$}	&
    \multicolumn{1}{l}{$\sum_{i,a} Z_a \left\langle \delta(\br_{ia})\right\rangle_\text{rg}$}  &
    \multicolumn{1}{l}{$\sum_{i,j} \left\langle \delta(\br_{ij})\right\rangle_\text{rg}$} &
    \multicolumn{1}{l}{$\langle H^{(2)}_\text{BP}\rangle$} \\
    \hline\\[-0.35cm]
    \multicolumn{3}{l}{Direct} \\
    $1.7\cdot 10^{-7}$  & $-$1.65               & 0.918	              &	0.016 9	            &	$-$0.205	\\
    $1.6\cdot 10^{-8}$  & $-$1.65	            & 0.919	              &	0.016 8	            &	$-$0.205	\\
    $1.4\cdot 10^{-9}$  & $-$1.65	            & 0.919	              &	0.016 78            &	$-$0.205 5	\\
    $3.8\cdot 10^{-10}$ & $-$1.654              & 0.919	              &	0.016 77            &	$-$0.205 5	\\
    \hline\\[-0.35cm]
    \multicolumn{3}{l}{Numerical drachmannization} \\
    $1.7\cdot 10^{-7}$  & $-$1.655	            & 0.919 33	          &	0.016 742           &	$-$0.206	\\
    $1.6\cdot 10^{-8}$  & $-$1.654 8	        & 0.919 335 5	      &	0.016 743 1         &	$-$0.205 8	\\
    $1.4\cdot 10^{-9}$  & $-$1.654 80	        & 0.919 335 7	      &	0.016 743 21        &	$-$0.205 75 \\
    $3.8\cdot 10^{-10}$ & $-$1.654 79	        & 0.919 335 8	      &	0.016 743 24        &	$-$0.205 733 \\
    \hline\\[-0.35cm]
    \multicolumn{3}{l}{Ref.~\cite{PuKoPa17}} \\    
    `direct' %
                    & $-$1.654 837 5(1)	    & 0.919 37(7)	      &	0.016 743 5(4)	    &	$-$0.205 682 (8)  \\
    `st. reg.' %
                    & $-$1.654 745 (2)	    & 0.919 336 211 2(18) &	0.016 743 278 1(7)	&	$-$0.205 689 (2)  \\
     `rECG \& m.reg.' %
                    & $-$1.654 744 522 5(1)	& 0.919 336 211(2)	  &	0.016 743 278 3(3)	&	$-$0.205 688 526(7)  \\
    \hline\hline\\[-0.35cm]
  \end{tabular}
  \begin{flushleft}
      Ref.~\cite{PuKoPa17}: $E_\text{nr}=-$1.174 475 714 220 443 4(5) $\Eh$.
  \end{flushleft}
\end{table}

\begin{table}[h]
  \caption{
    Molecular systems: numerically drachmannized expectation values and the leading-order relativistic correction. $-\delta E_\text{nr}$ is the estimated convergence error, in $\Eh$ of the (variational) non-relativistic energy. All converged and 1-2 additional digits are shown. All computations have been carried out with the same $(M,a,b)=(200,-31,31)$ parameterization of the Gaussian expansion.
    Convergence tables are provided as \som. 
    \label{tab:molecules} 
  }
\scalebox{0.95}{%
  \begin{tabular}{@{}l ll l l l l@{}}
    \hline\hline\\[-0.40cm]
    \multicolumn{1}{l}{} & 
    \multicolumn{1}{l}{$-\delta E_\text{nr}$} & 
    \multicolumn{1}{l}{$-\frac{1}{8}\sum_i \langle \bos{\nabla}_i^2 \bos{\nabla}_i^2 \rangle_{\text{rg}}$}  &
    \multicolumn{1}{l}{$\sum_{i,a} Z_a \langle \delta(\br_{ia}) \rangle_\text{rg}$}  &
    \multicolumn{1}{l}{$\sum_{i,j} \langle \delta(\br_{ij}) \rangle_\text{rg}$} &
    \multicolumn{1}{l}{$\langle H_\text{OO}\rangle$} &
    \multicolumn{1}{l}{$\langle H^{(2)}_\text{BP}\rangle$}
    \\
    \hline\\[-0.35cm]
    $\text{H}_2^{+}$ %
    & $10^{-12}$$^\text{b}$
    & $-$0.624 952 638 6 
    &    0.318 293 258 6 
    & -- 
    & -- 
    & $-$0.124 978 757 0 \\
    b \cite{HoKe90}$^\text{c}$ %
    & $10^{-9}$ 
    & $-$0.624 952 7
    & 0.318 293 2
    & -- 
    & -- 
    & $-$0.124 978 9 \\
    \hline\\[-0.35cm]
    $\text{H}_3^{+}$$^\text{c}$ %
    &  $10^{-9}$
    &  $-$1.933 449 7
    &     1.089 654 23
    &     0.018 334 666 
    &  $-$0.057 217 520  
    &  $-$0.221 442 3 \\
    i \cite{JeIrFeMa22}$^\text{c}$ %
    &  $10^{-9}$
    &  $-$1.933 424         
    &     1.089 655         
    &     0.018 335       
    &  $-$0.057 218 
    &  $-$0.221 422 \\
    \hline\\[-0.35cm]
    $\text{HeH}^{+}$ %
    &  $10^{-9}$
    &  $-$13.419 497 
    &      7.216 244 8
    &      0.101 121 71
    &   $-$0.141 241 61  
    &   $-$1.907 804  \\
    i \cite{JeFeMa22}$^\text{c}$ %
    &  $10^{-9}$
    &  $-$13.419 29       
    &      7.216 25         
    &      0.101 122
    &   $-$0.141 242
    &   $-$1.907 58 \\
    \hline\\[-0.35cm]
    $\text{He}_2^{+}$ %
    & $10^{-8}$
    & $-$23.867 58
    &    12.537 54
    &     0.117 596 4
    &  $-$0.086 577
    &  $-$3.890 79 \\
    i \cite{FeKoMa20}$^\text{c}$ %
    & $10^{-8}$
    & $-$23.866 14
    &    12.537 51
    &     0.117 63
    &  $-$0.086 58
    &  $-$3.889 30 \\
    \hline\\[-0.35cm]
    $\text{H}_3$  %
    & $10^{-8}$
    & $-$2.277 870 
    &    1.236 566 
    &    0.016 712 
    & $-$0.046 780 1
    & $-$0.329 754 2
    \\
    \hline\\[-0.35cm]
    $\text{He}_2$ %
    & $10^{-6}$
    & $-$24.157 271
    &    12.660 991
    &     0.119 967 0
    &  $-$0.073 154 
    &  $-$3.965 700 
    \\
    \hline\hline
  \end{tabular}
}
  \begin{flushleft}
    $^\text{a}$ %
    The electronic state and molecular geometry was
    H$_2^+$, $\tilde{\text{A}}\ ^2\Sup$: $R_\text{p-p}=12.00$~bohr; 
    H$_2$, $\text{X}\ ^1\Sgp$: $R_\text{p-p}=1.40$~bohr; 
    H$_3^+$, $^1\text{A}_1'$: $R_\text{p-p}=1.65$~bohr, $\vartheta_{\text{p}_1-\text{p}_2-\text{p}_3}=60^\text{o}$;
    H$_3$, $1\ ^2\Sigma^+$: 
    $R_{\text{p}_1-\text{p}_2}=1.4015$~bohr, 
    $R_{\text{p}_2-\text{p}_3}=5.8113$~bohr,
    $\vartheta_{\text{p}_1-\text{p}_2-\text{p}_3}=180^\text{o}$; 
    HeH$^+$, $1\ ^1\Sigma^+$: 
    $R_{\text{p}-\alpha}=1.46$~bohr;
    He$_2^+$, $1\ ^2\Sup$: 
    $R_{\alpha-\alpha}=2.042$~bohr, 
    He$_2$, $a(1)\ ^3\Sup$: 
    $R_{\alpha-\alpha}=2.00$~bohr. \\
    $^\text{b}$~The non-relativistic energy has been reported with higher precision in Ref.~\cite{BeMe16}. \\
    $^\text{c}$~Earlier available computation with 
    regularization type: integral transform (i) \cite{PaCeKo05,JeIrFeMa22}, special basis set (b) \cite{HoKe90}.
  \end{flushleft}
\end{table}

This work reports a robust and generally applicable numerical drachmannization approach for the computation of the leading-order relativistic corrections with floating explicitly correlated Gaussian basis functions used to describe small molecule electronic structure to high precision.
Formerly incalculable $\langle 1/r_{ix} 1/ r_{jy}\rangle$-type floating ECG integrals are made calculable using a simple and robust Gaussian approximation to one of the $1/r$ factors. The computational approach is demonstrated to be numerically robust and generally applicable for two-, three-, and four-electron atomic, diatomic, and polyatomic systems and along potential energy curves (and potentially, surfaces). 
The numerical drachmannization approach developed in this work opens the route to the automated computation of high-precision relativistic corrections to potential energy curves and surfaces in the future.

\vspace{0.5cm}
\begin{acknowledgments}
\noindent Financial support of the European Research Council through a Starting Grant (No.~851421) is gratefully acknowledged. We thank Péter Jeszenszki and Robbie Ireland for discussions and joint work over the past years. 
\\
\end{acknowledgments}

\clearpage
\providecommand{\noopsort}[1]{}\providecommand{\singleletter}[1]{#1}%
%

\clearpage

~\\[0.cm]
\begin{center}
\begin{minipage}{1.\linewidth}
\centering
\textbf{Supplementary Material} \\[0.5cm]

\textbf{%
Regularized relativistic corrections for polyelectronic and polyatomic molecules with explicitly correlated Gaussians \\[0.5cm]
}
\end{minipage}
~\\[0.5cm]
\begin{minipage}{0.6\linewidth}
\centering

Balázs Rácsai,$^{1}$ Dávid Ferenc,$^1$ Ádám Margócsy,$^1$ and Edit M\'atyus$^{1,\ast}$ \\[0.15cm]

$^1$~\emph{ELTE, Eötvös Loránd University, Institute of Chemistry, 
Pázmány Péter sétány 1/A, Budapest, H-1117, Hungary} \\[0.15cm]
$^\ast$ edit.matyus@ttk.elte.hu \\
\end{minipage}
~\\[0.15cm]
(Dated: \today)
\end{center}



During the integral calculations, the following notation is used with 
$\bos{A}_\mu,\bos{A}_\nu \in\mathbb{R}^{n\times n}$, 
$\bos{s}_\mu,\bos{s}_\nu \in\mathbb{R}^{3n}$, 
and $\bos{I}_3$ is the $3\times 3$ unit matrix,
\begin{align}
  \bos{\underline{A}}_{\mu} 
  &= 
  \bos{A}_{\mu} \otimes \bos{I}_{3} \; ,
  \quad
  \bos{A}_{\mu\nu} 
  = 
  \bos{A}_{\mu} + \bos{A}_{\nu} \; ,
  \quad
  \be = \bos{\underline{A}}_{\mu}\bs_{\mu} + \bos{\underline{A}}_{\nu}\bs_{\nu} \; , \nonumber \\
  \eta &= \bos{s}_{\nu}^\text{T} \bos{\underline{A}}_{\nu}\bos{s}_{\nu} + \bos{s}_{\mu}^\text{T} \bos{\underline{A}}_{\mu}\bos{s}_{\mu} \; , 
  \quad
  \gamma = \bos{e}^\text{T} \bos{\underline{A}}_{\mu\nu}^{-1}\bos{e} - \eta  \; , 
  \quad
  \bos{s} = \bos{\underline{A}}_{\mu\nu}^{-1}\bos{e} \; ,  
  \quad
  S_{\mu\nu} = \eem^\gamma \frac{\pi^{\frac{3n}{2}}}{|\bos{A}_{\mu\nu}|^{\frac{3}{2}}} \; ,
\end{align}
where the $\mu, \nu$ indices are suppressed in $\bos{e}, \eta,$ and $\gamma$ for brevity. The short notation $\underline{\bos{M}}=\bos{M}\otimes \bos{I}_3$ will be used.

\subsection{$\bra{\varphi_{\mu}}\frac{1}{r_{ij}}\frac{1}{r_{pq}}\ket{\varphi_{\nu}}$-type integrals}
We start with the product of two $1/r$ electron-electron interaction terms ($i,j,p,q=1,2,\ldots,n$ label electronic coordinates), and use the Gaussian approximation for one of the factors, 
\begin{align}
  \bra{\varphi_\mu}\frac{1}{r_{ij}}\frac{1}{r_{pq}}\ket{\varphi_\nu} 
  &\cong 
  \sum_{m=1}^M 
    w_m 
    \bra{\varphi_\mu}
      \frac{1}{r_{ij}}\, \eem^{-p_m r_{pq}^2}
    \ket{\varphi_\nu} \; .
   \label{eq:approx-begin}
  \\
  &= 
  \sum_{m=1}^M w_m\; %
  \int_{\mathbb{R}^{3n}} \dd\br\
    \frac{1}{r_{ij}}\, \eem^{-p_m r_{pq}^2}
    \eem^{-\eta}\eem^{-\br^\text{T} \ubA_{\mu\nu} \br + 2\;\bos{e}^\text{T}\br} \; ,
\end{align}
where we used the relation for the product of two fECGs, 
\begin{align}
  \eem^{-(\br-\bs_\mu)^T\ubA_\mu(\br-\bs_\mu)}
  \eem^{-(\br-\bs_\nu)^T\ubA_\nu(\br-\bs_\nu)}
  =
  \eem^{-\eta} \eem^{-\br^\text{T} \ubA_{\mu\nu} \br + 2\;\bos{e}^\text{T}\br} \; .
  \label{eq:fecg-contraction}
\end{align}
Then, we proceed by using 
\begin{align}
    \frac{1}{r_{ij}} 
    = 
    \frac{2}{\sqrt{\pi}} \int_{0}^{\infty}\dd t \, \eem^{-r_{ij}^2 \, t^2} \; ,
    \label{eq:1d-gauss}
\end{align}
and thus, 
\begin{align}
  \bra{\varphi_\mu}\frac{1}{r_{ij}}\frac{1}{r_{pq}}\ket{\varphi_\nu} 
  &= 
  \sum_{m=1} ^M w_m\; %
    \frac{2}{\sqrt{\pi}} \eem^{-\eta}  
  \int_{0}^{\infty}\dd t 
  \int_{\mathbb{R}^{3n}} \dd\br \,
    \eem^{-p_m r_{pq}^2 - t^2 r_{ij}^2}
    \eem^{-\br^\text{T} \ubA_{\mu\nu} \br + 2\;\bos{e}^\text{T} \br} \; .
        \label{eq:ueeuee-start}
\end{align}
Next, we introduce a matrix $\bos{J}\in\mathbb{R}^{n\times n}$ defined as 
\begin{align}
  &(\bos{J}_{ij})_{\alpha \beta} 
  = 
  \delta_{\alpha\beta}(\delta_{\alpha i}+\delta_{\alpha j}) 
  - 
  (\delta_{\alpha i}\delta_{\beta j} + \delta_{ \beta i}\delta_{\alpha j})(1 - \delta_{\alpha\beta}) \; ,
  \quad\quad
  \alpha, \beta=1,\ldots,n. 
  \label{eq:Jdef}
\end{align}
Then, we can write
\begin{align}
    r^2_{ij} &= \br^\text{T} \ubJ_{ij} \br \; , \\
    r^2_{pq} &= \br^\text{T} \ubJ_{pq} \br \; , 
\end{align}
and 
\begin{align}
    \bra{\varphi_\mu}\frac{1}{r_{ij}}\frac{1}{r_{pq}}\ket{\varphi_\nu} 
    = 
    \sum_{m=1} ^M w_m\;\frac{2}{\sqrt{\pi}}\;\eem^{-\eta} 
    \int_{0}^{\infty} \dd t \int_{\mathbb{R}^{3n}} \dd \br  \, 
    \eem^{-\br^\text{T}(\ubA_{\mu\nu} + p_m\ubJ_{pq} + t^2\ubJ_{ij})\br + 2\,\bos{e}^\text{T}\br } \; .
    \label{eq:ueeuee-basis}
\end{align}
By defining $\ubA'_{\mu\nu} = \ubA_{\mu\nu} + p_m\ubJ_{pq}$, which is positive definite for all $p_m$ used in the Gaussian expansion, and using the identity for positive definite $\bos{M}\in\mathbb{R}^{3n\times 3n}$
\begin{align}
  \int_{\mathbb{R}^{3n}} 
   \dd \bos{x}\  
     \eem^{-\bos{x}^\text{T}\bos{M}\bos{x} + \bos{y}^\text{T}\bos{x}}
    = 
    \frac{\pi^{\frac{n}{2}}}{|\bos{M}|^{\frac{1}{2}}}\;%
    \eem^{\frac{1}{4} \bos{y}^\text{T}\bos{M}^{-1}\bos{y} } \; ,
\end{align}
we obtain
\begin{align}
  \bra{\varphi_\mu}\frac{1}{r_{ij}}\frac{1}{r_{pq}}\ket{\varphi_\nu} 
  = 
  \sum_{m=1}^M w_m\;
  \frac{2}{\sqrt{\pi}}\,\pi^{\frac{3n}{2}}\;\eem^{-\eta}
  \int_{0}^{\infty} \dd  t \, |\bA'_{\mu\nu} + t^2\bJ_{ij}|^{-\frac{3}{2}}
  \eem^{\bos{e}^\text{T}\left(\ubA'_{\mu\nu} + t^2\ubJ_{ij}\right)^{-1}\bos{e}} 
  \; .
     \label{eq:ueeuee-deriv1}
\end{align}
Since both the $\bJ_{ij}$ and $\bJ_{pq}$ matrices are rank one, we can use the known mathematical relations for a non-singular $\bos{A}\in\mathbb{R}^{n\times n}$ and a rank-one $\bos{B}\in\mathbb{R}^{n\times n}$ matrix, \emph{e.g.,} Ref.~\cite{SuVaBook98,StPaAd16},
\begin{align}
  |\bos{I} + a\bos{B}| 
  &= 
  1 + a\; \mathrm{Tr}\bos{B} \; ,\label{eq:rankone1}\\
  (\bA + \bos{B})^{-1} 
  &= 
  \bA^{-1} - \frac{\bA^{-1}\bos{B}\bA^{-1}}{1 + \mathrm{Tr}(\bos{B}\bA^{-1})} \; , \label{eq:rankone2}
\end{align}
and thus, the integrand can be further simplified to
\begin{align}
  \bra{\varphi_\mu}\frac{1}{r_{ij}}\frac{1}{r_{pq}}\ket{\varphi_\nu} 
  &= 
  \sum_{m=1}^M w_m\;
    \frac{2}{\sqrt{\pi}}\;
    \pi^{\frac{3n}{2}}\;
    |\bA'_{\mu\nu}|^{-\frac{3}{2}}
    \eem^{\bos{e}^\text{T}\ubA'^{-1}_{\mu\nu}\bos{e}-\eta}  \;\nonumber \\
    &
    \int_{0}^{\infty} \dd t\ 
    \frac{%
      1
    }{%
      (1 + t^2\mathrm{Tr}[\bA'^{-1}_{\mu\nu}\bJ_{ij}])^{\frac{3}{2}}
    }\ 
    \eem^{%
      -\frac{t^2}{1+t^2\mathrm{Tr}(\bA'^{-1}_{\mu\nu}\bos{J}_{ij})}
      \bos{e}^\text{T}\ubA'^{-1}_{\mu\nu}\ubJ_{ij}\ubA'^{-1}_{\mu\nu}\bos{e}
    } \; .
    \label{eq:ueeuee-mess}
\end{align}
For convenience, we introduce three temporary quantities,
\begin{align}
    S'_{\mu\nu} 
    &= 
    \pi^{\frac{3n}{2}}\; |\bA'_{\mu\nu}|^{-\frac{3}{2}}
    \eem^{\bos{e}^\text{T}\ubA'^{-1}_{\mu\nu}\bos{e}-\eta}\; ,
    \label{eq:temp1}
    \\
    a' &= \mathrm{Tr}(\bA'^{-1}_{\mu\nu}\bJ_{ij}) \; ,
    \label{eq:temp2}
    \\
    \beta' &= \bos{e}^\text{T} \ubA'^{-1}_{\mu\nu}\ubJ_{ij}\ubA'^{-1}_{\mu\nu} \bos{e}  \; , 
    \label{eq:temp3}
\end{align}
which allows us to write the result in a compact form, 
\begin{align}
  \bra{\varphi_\mu}\frac{1}{r_{ij}}\frac{1}{r_{pq}}\ket{\varphi_\nu} 
  =
  \sum_{m=1}^M w_m\;
    \frac{2}{\sqrt{\pi}}\; S'_{\mu\nu}
  \int_{0}^{\infty} \dd t \; 
  (1 + t^2a')^{-\frac{3}{2}}
  \eem^{\frac{-t^2\beta'}{1+t^2a'}} \; .
\end{align}
Finally, by the substitution $z = \frac{t^2\beta'}{1+t^2a'}$, we can use the definition of the error function to arrive at
\begin{align}
  \bra{\varphi_\mu}\frac{1}{r_{ij}}\frac{1}{r_{pq}}\ket{\varphi_\nu} 
  \cong 
  \sum_{m=1}^M w_m\; \frac{S'_{\mu\nu}}{\sqrt{\beta'}}\;
  \mathrm{erf}\left[\sqrt{\frac{\beta'}{a'}}\right]
  \; . \label{eq:ueeuee-niceresult}
\end{align}
This result is similar to the well-known expression for the  $\bra{\varphi_\mu}\frac{1}{r_{ij}}\ket{\varphi_\nu}$ matrix element. For a computer implementation, we express the temporary variables, Eqs.~\eqref{eq:temp1}--\eqref{eq:temp3}, in terms of the original variables through the repeated use of the identities of Eqs.~\eqref{eq:rankone1} and \eqref{eq:rankone2}.
\begin{align}
  S'_{\mu\nu} 
  &= 
  S_{\mu\nu} \left(1 + t_2\right)^{-\frac{3}{2}}
  \eem^{-\frac{p_m}{1 + p_m t_2} \sum_{\alpha=1}^3 (s_{p \alpha} - s_{q \alpha})^2}
  \; , \\
  a' 
  &= 
  \frac{t_1 + p_m(t_1 t_2 - t_3)}{1 + p_m t_2} 
  \; , \\
  \beta' 
  &= 
  \frac{1}{(1+p_m t_2)^2}
  \sum_{\alpha=1}^3 
    \lbrace%
      (s_{i \alpha} - s_{j \alpha})
      + 
      p_m 
      [%
      t_2 (s_{i \alpha} - s_{j \alpha}) 
      - 
      \sqrt{t_3} (s_{p \alpha} - s_{q \alpha})
      ]
    \rbrace^2 \; 
\end{align}
with 
\begin{align}
    t_1 &= \mathrm{Tr}[\bA^{-1}_{\mu\nu}\bJ_{ij}]  \; , \\
    t_2 &= \mathrm{Tr}[\bA^{-1}_{\mu\nu}\bJ_{pq}] \; , \\
    t_3 &= \mathrm{Tr}[\bA^{-1}_{\mu\nu}\bJ_{pq}\bA^{-1}_{\mu\nu}\bJ_{ij}]  \; .
  \label{eq:eeeeparams1}
\end{align}
\noindent %
After substitutions, we arrive at the final, complete expression,
\begin{align}
  \bra{\varphi_\mu}\frac{1}{r_{ij}}\frac{1}{r_{pq}}\ket{\varphi_\nu} 
  &= 
  S_{\mu\nu} 
  \sum_{m=1}^M 
    w_m\ \eem^{x_{m,1}}\ \frac{1}{x_{m,2}} \erf[x_{m,3}]
  \label{eq:finalinteeee}
\end{align}
with
\begin{align}
  x_{m,1} 
  &= 
  -\frac{p_m}{1 + p_m t_2} 
  \sum_{\alpha=1}^3 
    (s_{p \alpha} - s_{q \alpha})^2 
  \; , \nonumber \\
  x_{m,2} 
  &= 
  \left[
  (1 + p_m t_2) 
  \sum_{\alpha=1}^3 
  \lbrace %
    (s_{i \alpha} - s_{j \alpha})
    + 
    p_m 
    [%
    t_2 (s_{i \alpha} - s_{j \alpha}) 
    - 
    \sqrt{t_3} (s_{p \alpha} - s_{q \alpha})
    ]
  \rbrace^2   
  \right]^{\frac{1}{2}}
  \; , \nonumber \\
  x_{m,3} 
  &= 
  \left[%
  \frac{%
    \sum_{\alpha=1}^3 
      \lbrace%
        (s_{i \alpha} - s_{j \alpha})
         + 
         p_m 
         [%
         t_2 (s_{i \alpha} - s_{j \alpha}) 
         - 
         \sqrt{t_3} (s_{p \alpha} - s_{q \alpha})
         ]
      \rbrace^2
  }{%
    (1 + p_m t_2)[t_1 + p_m (t_1 t_2 - t_3)]
  } 
  \right]^{\frac{1}{2}}
  \; .
  \label{eq:eeeeparams12}
\end{align}
For a numerically stable implementation, the explicit squares in the expressions are useful to ensure that a non-negative value is obtained under the square root (even for small numerical quantities in finite precision arithmetic).

With this result, we can, in principle, evaluate all matrix elements containing two $1/r$ electron-electron interaction factors, however, better numerical stability can be obtained if special cases are treated separately.

\paragraph{Zero-shift case}
First, if the $\bos{s}_\mu,\bos{s}_\nu$ shift vectors are zero, and thus, $\be=\bs=0$, it is better to evaluate the mathematically simplified expression.
According to our experiments, the $p_m$ approximation parameter tends to magnify small values, making the computation numerically unstable.
So, for $\bos{s}=0$, it is better to use the simplified expression,
\begin{align}
  \bra{\varphi_\mu}\frac{1}{r_{ij}}\frac{1}{r_{pq}}\ket{\varphi_\nu}
  =& \; 
  \frac{2}{\sqrt{\pi}} S_{\mu\nu}  
  \sum_{m=1}^M w_m  \frac{1}{(1 + p_m t_2)\sqrt{t_1 + p_m (t_1t_2-t_3)}} \; .
  \label{eq:ueeuee-case2}
\end{align}
Alternatively, we could also use the already available analytic expressions (\emph{e.g.,} calculated for atoms in Ref.~\cite{IrJeMaMaRoPo21}), which served as an important test case for the present calculations and implementation.
A smooth connection of the zero and the non-zero shift cases can be realized by considering a Taylor expansion of the full expression for small $\bos{s}$ values.

\paragraph{Identical factors}
The other important special case, which is best treated separately corresponds to identical $1/r$ factors, \emph{i.,e.,} $\frac{1}{r_{ij}^2}$. The best numerical stability can be obtained, if the general expression is simplified manually according to $\bJ_{ij}=\bJ_{pq}$, resulting in 
\begin{align}
  \bra{\varphi_\mu}\frac{1}{r_{ij}^2}\ket{\varphi_\nu} 
  =
  S_{\mu\nu} 
  \sum_{m=1}^M w_m 
    \eem^{-\frac{p_m}{1 + p_m t_1}\kappa_1}
    \frac{1}{\sqrt{\kappa_1(1 + p_m t_1)}}
    \erf\left[%
      \sqrt{\frac{\kappa_1}{t_1(1 + p_m t_1)}}
    \right] \; 
    \label{eq:ueeuee-case3} 
\end{align}
with
\begin{align}
  \kappa_1 
  = 
  \bos{s}\ubJ_{ij}\bos{s} 
  =
  \sum_{\alpha=1}^3 (s_{i \alpha} - s_{j \alpha})^2 \; . 
\end{align}
The zero-shift emerges as a special sub-case: 
\begin{align}
  \bra{\varphi_\mu}\frac{1}{r_{ij}^2}\ket{\varphi_\nu} 
  = 
  \frac{2}{\sqrt{\pi}} S_{\mu\nu} 
  \sum_{m=1}^M 
    w_m \frac{1}{(1 + p_m t_1)\sqrt{t_1}} \; .
  \label{eq:ueeuee-case4}
\end{align}

\subsubsection{%
  The $\bra{\varphi_{\mu}}\frac{1}{r_{jk}}\frac{1}{|r_i - R_a|}\ket{\varphi_{\nu}}$ Matrix Element
}
Next, we consider the fECG matrix element for the product of an electron-electron and an electron-nucleus interaction factor, 
\begin{align}
  \bra{\varphi_\mu}\frac{1}{r_{jk}}\frac{1}{|\br_i - \bR_a|}\ket{\varphi_\nu} 
  \cong 
  \sum_{m=1}^M w_m 
    \bra{\varphi_\mu}\frac{1}{|\br_i - \bR_a|}\, \eem^{-p_mr_{jk}^2} \ket{\varphi_\nu} \; ,
\end{align}
and along similar steps as in the previous case, we arrive at
\begin{align}
  \bra{\varphi_\mu}\frac{1}{r_{jk}}\frac{1}{|\br_i - \bR_a|}\ket{\varphi_\nu} 
  &= 
  \sum_{m=1}^M w_m\;
    \frac{2}{\sqrt{\pi}} \eem^{-\eta}  
    \int_{0}^{\infty}\dd t 
    \int_{\mathbb{R}^{3n}} \dd\br\,
      \eem^{-p_m r_{jk}^2 - t^2 |\br_i - \bR_a|^2 } 
      \eem^{-\br^\text{T} \ubA_{\mu\nu} r + 2\;\bos{e}^\text{T}\br } \; .
\end{align}
We rewrite
\begin{align}
  |\br_i - \bR_a|^2 
  &= 
  (\br-\ba)^T\ubE_{ii}(\br-\ba) 
  \; 
\end{align}
with the definition 
\begin{align}
  (\ba)_{l\alpha}= R_{a,\alpha}
  \quad\quad\text{and}\quad\quad
  (\bos{E}_{ii})_{ll} = \delta_{il}\; , \quad l=1,\ldots,n\;,\alpha=1(x),2(y),3(z) \; .
\end{align}
Then, 
\begin{align}
  r_{jk}^2 
  &= 
  \br^\text{T}\ubJ_{jk}\br = (\br-\ba)^\text{T}\ubJ_{jk}(\br-\ba) \; ,
\end{align}
where the last step is a mathematical identity with the original definition of the $\bJ_{jk}$ matrix, Eq.~\eqref{eq:Jdef}. 
Then, further rearrangements
\begin{align}
  \eem^{-\eta}
  \eem^{-\br^\text{T}\ubA_{\mu\nu}\br + 2\be^\text{T}\br}
  &= 
  \eem^{\gamma}
  \eem^{-(\bs-\ba)^\text{T}\ubA_{\mu\nu}(\bs-\ba) - (\br-\ba)^\text{T}\ubA_{\mu\nu}(\br-\ba)}
  \eem^{2(\bs-\ba)^\text{T}\ubA_{\mu\nu}(\br-\ba)} 
  \; ,
    \label{eq:rAr-expand}
\end{align}
and the $\bos{\xi} = \br - \ba$ coordinate transformation (leaving the integration boundaries unchanged) lead to
\begin{align}
  &\bra{\varphi_\mu}\frac{1}{r_{jk}}\frac{1}{|\br_i - \bR_a|}\ket{\varphi_\nu} 
  =
  \nonumber \\
  &\quad\quad\quad\quad
  \frac{2}{\sqrt{\pi}} 
  \sum_{m=1}^M w_m\, 
  \eem^{\gamma - (\bs - \bos{a})^\text{T} \ubA_{\mu\nu}(\bs - \bos{a})} 
  \int_{\mathbb{R}^{3n}} \dd\bos{\xi} \; 
  \int_0^{\infty} \dd t\  
    \eem^{-\bos{\xi}^\text{T}(\ubA_{\mu\nu} + p_m \ubJ_{jk} + t^2\ubE_{ii})\bos{\xi} + 2{\bos{e}'}^\text{T}\bos{\xi}} \; ,
    \label{eq:ueeuen-finjump}
\end{align}
where $\bos{e}' = \ubA_{\mu\nu}(\bs - \bos{a})$. 
Then, the calculation proceeds similarly to the electron-electron case in the previous section, and the final result is ($r_{ia}=|\br_i-\bos{R}_a|$)
\begin{align}
  \bra{\varphi_\mu}\frac{1}{r_{jk}}\frac{1}{r_{ia}}\ket{\varphi_\nu} 
  &= 
  S_{\mu\nu} 
  \sum_{m=1}^M 
  w_m\ \eem^{y_{m,1}}\ \frac{1}{y_{m,2}} \erf[y_{m,3}]
\end{align}
with
\begin{align}
  y_{m,1} 
  &= 
  -\frac{p_m}{1 + p_m u_2} 
  \sum_{\alpha=1}^3 
    (s_{j \alpha} - s_{k \alpha})^2 
  \nonumber\\
  y_{m,2} 
  &= 
  \left[%
  (1 + p_m u_2) 
  \sum_{\alpha=1}^3 
  \lbrace%
    (s_{i \alpha} - a_{i \alpha})
    + 
    p_m 
    [%
    u_2 (s_{i \alpha} - a_{i \alpha}) 
    - 
    \sqrt{u_3} (s_{j \alpha} - s_{k \alpha})
    ]
  \rbrace^2   
  \right]^\frac{1}{2}
  \nonumber\\
  y_{m,3} 
  &= 
  \left[%
  \frac{%
    \sum_{\alpha=1}^3 
      \lbrace%
        (s_{i \alpha} - {a}_{i \alpha})
         + 
         p_m 
         [%
         u_2 (s_{i \alpha} - {a}_{i \alpha}) 
         - 
         \sqrt{u_3} (s_{j \alpha} - s_{k \alpha})
         ]
      \rbrace^2
  }{%
    (1 + p_m u_2)[ u_1 + p_m (u_1 u_2 - u_3) ]
  } 
  \right]^\frac{1}{2}
  \; 
  \label{eq:eeen}
\end{align}
and 
\begin{align}
  u_1 =
  &\;
  \mathrm{Tr}[ \bA^{-1}_{\mu\nu}\bE_{ii} ] \label{eq:ueeuen-t1} \; , \\
  u_2
  =&\;
  \mathrm{Tr}[ \bA^{-1}_{\mu\nu}\bJ_{jk} ] \; , \\
  u_3
  =&\;
  \mathrm{Tr}[ \bA^{-1}_{\mu\nu}\bJ_{jk}\bA^{-1}_{\mu\nu}\bE_{ii} ] \; .
\end{align}
The zero-shift case of Eq.~\eqref{eq:ueeuee-case2} also has been considered separately.

\subsubsection{The $\bra{\varphi_{\mu}}\frac{1}{|r_i -R_a|}\frac{1}{|r_j -R_b|}\ket{\varphi_{\nu}}$ Matrix Element}
Finally, we calculate the integral containing two $1/r$ electron-nucleus interaction factors using the Gaussian approximation for one of the factors,
\begin{align}
  \bra{\varphi_\mu} \frac{1}{|\br_i - \bR_a|} \frac{1}{|\br_j - \bR_b|} \ket{\varphi_\nu} 
  \cong 
  \sum_{m=1}^M w_m 
    \bra{\varphi_\mu}\frac{1}{|\br_i - \bR_a|}\, \eem^{-p_m |\br_j - \bR_b|^2} \ket{\varphi_\nu} \; .
\end{align}
The general derivation is almost completely identical to the previous ones, but we first rewrite
\begin{align}
    |\br_i - \bR_a|^2 &= (\br-\bc)^\text{T}\ubE_{ii}(\br-\bc) \; ,\\
    |\br_j - \bR_b|^2 &= (\br-\bc)^\text{T}\ubE_{jj}(\br-\bc) \; ,
\end{align}
and define the $\bc$ vector as ($i\neq j$, regarding the $i=j,a\neq b$ case, please see below) 
\begin{align}
  c_{k\alpha}
  =
  \delta_{ki} {R}_{a\alpha}
  +
  \delta_{kj} {R}_{b\alpha}
  \quad 
  k=1,\ldots,n; \alpha=1(x),2(y),3(z).
  \label{eq:cdef}
\end{align}
After substitutions, we arrive at the final, complete expression, 
\begin{align}
  \bra{\varphi_\mu}\frac{1}{r_{ia}}\frac{1}{r_{jb}}\ket{\varphi_\nu} 
  &= 
  S_{\mu\nu} 
  \sum_{m=1}^M 
    w_m\ \eem^{z_{m,1}}\ \frac{1}{z_{m,2}} \erf[z_{m,3}]
  \label{eq:finalintenen}
\end{align}
with
\begin{align}
  z_{m,1} 
  &= 
  -\frac{p_m}{1 + p_m v_2} 
  \sum_{\alpha=1}^3 
    (s_{j \alpha} - c_{j \alpha})^2 
  \; , \nonumber \\
  z_{m,2} 
  &= 
  \left[%
  (1 + p_m v_2) 
  \sum_{\alpha=1}^3 
  \lbrace %
    (s_{i \alpha} - c_{i \alpha})
    + 
    p_m 
    [%
    v_2 (s_{i \alpha} - c_{i \alpha}) 
    - 
    \sqrt{v_3} (s_{j \alpha} - c_{j \alpha})
    ]
  \rbrace^2   
  \right]^\frac{1}{2}
  \; , \nonumber \\
  z_{m,3} 
  &= 
  \left[%
  \frac{%
    \sum_{\alpha=1}^3 
      \lbrace%
        (s_{i \alpha} - c_{i \alpha})
         + 
         p_m 
         [%
         v_2 (s_{i \alpha} - c_{i \alpha}) 
         - 
         \sqrt{v_3} (s_{j \alpha} - c_{j \alpha})
         ]
      \rbrace^2
  }{%
    (1 + p_m v_2)[v_1 + p_m (v_1 v_2 - v_3)]
  } 
  \right]^\frac{1}{2}
  \; .
  \label{eq:enenparams12}
\end{align}
and 
\begin{align}
  v_1=&\;\mathrm{Tr}[\bA^{-1}_{\mu\nu}\bE_{ii}] \; , \label{eq:v1} \\
  v_2=&\;\mathrm{Tr}[\bA^{-1}_{\mu\nu}\bE_{jj}] \; , \label{eq:v2} \\
  v_3=&\;\mathrm{Tr}[\bA^{-1}_{\mu\nu}\bE_{jj}\bA^{-1}_{\mu\nu}\bE_{ii}] \; . 
  \label{eq:enenparams1}
\end{align}
The zero-shift cases and the $\frac{1}{|\br_i - \bR_a|^2}$ case have also been considered according to the earlier discussion.

\paragraph{Special case with $i=j$} The integral of $\frac{1}{|\br_i - \bR_a|}\frac{1}{|\br_i - \bR_b|}$ requires further attention. Then, the $\bc$ vector, Eq.~\eqref{eq:cdef}, does not account for the $i=j$ case for arbitrary $a$ and $b$ indices.
In this case, we start with
\begin{align}
  &\bra{\varphi_\mu} \frac{1}{|\br_i - \bR_a|} \frac{1}{|\br_i - \bR_b|} \ket{\varphi_\nu} 
  \nonumber \\
  &\quad = 
  \sum_{m=1}^M 
    w_m \frac{2}{\sqrt{\pi}}
    \eem^{-\eta}
    \int_{\mathbb{R}^{3n}} \dd\br
    \int_{0}^{\infty} \dd t \;  
      \eem^{-t^2|\br_i-\bR_a|^2}
      \eem^{-p_m |\br_i-\bR_b|^2}
      \eem^{-\br^\text{T} \ubA_{\mu\nu} \br + 2\be^\text{T} \br} \; .
\end{align}
and define 
\begin{align}
  a_{k\alpha} = \delta_{ki} R_{a\alpha} 
  \quad\text{and}\quad
  b_{k\alpha} = \delta_{kj} R_{b\alpha} \;\quad k=1,\ldots,n\ ;\ \alpha=1(x),2(y),3(z).
\end{align}
and write
\begin{align}
  |\br_i-\bR_a|^2 
  &= 
  (\br-\ba)^\text{T} \ubE_{ii}(\br-\ba) \; , \\
  |\br_i-\bR_b|^2 
  &= 
  (\br-\bb)^\text{T} \ubE_{ii}(\br-\bb) \; \mathrm{.}
\end{align}
For further calculations, it will be useful to re-write 
\begin{align}
  (\br-\bb)^\text{T} \ubE_{ii}(\br-\bb)
  = 
  (\br-\ba)^\text{T} \ubE_{ii}(\br-\ba)
  -2(\br-\ba)^\text{T} \ubE_{ii} (\bb-\ba) 
  + 
  (\ba-\bb)^\text{T} \ubE_{ii} (\ba-\bb) \; ,
\end{align}
which is combined with the identity in Eq.~\eqref{eq:rAr-expand}, the $\bos{\xi}=(\br-\ba)$ substitution, and $(\be'')^\text{T}=(\bs-\ba)^\text{T}\ubA_{\mu\nu}+(\bb-\ba)^\text{T}p_m\ubE_{ii}$, to obtain 
\begin{align}
  &\bra{\varphi_\mu}\frac{1}{|\br_i - \bR_a|}\frac{1}{|\br_i - \bR_b|}\ket{\varphi_\nu} 
  \nonumber \\
  &\quad 
  = 
  \sum_{m=1}^M 
    w_m \frac{2}{\sqrt{\pi}}
    \eem^{\gamma-(\bs-\ba)^\text{T} \ubA_{\mu\nu}(\bs-\ba)-(\ba-\bb)^\text{T} p_m \ubE_{ii} (\ba-\bb)}
    \nonumber\\
  &\quad\quad\quad\quad
  \int_{\mathbb{R}^{3n}} \dd\bos{\xi}\int_{0}^\infty \dd t \;  
    \eem^{-\bos{\xi}^\text{T}(\ubA_{\mu\nu}+p_m\ubE_{ii}+t^2\ubE_{ii})\bos{\xi} + 2\,(\be'')^\text{T}\bos{\xi} } \; .
\end{align}
This integral can be calculated along analogous steps as Eq.~\eqref{eq:ueeuee-basis}.
The final result takes the form, with $v_1$ having been defined in Eq.~\eqref{eq:v1}, 
\begin{align}
  \bra{\varphi_\mu}&\frac{1}{|\br_i - \bR_a|}\frac{1}{|\br_i - \bR_b|}\ket{\varphi_\nu} 
  = 
  S_{\mu\nu} \sum_{m=1}^M w_m \eem^{\bar{z}_{m,1}} \frac{1}{{\bar{z}_{m,2}}} \erf[{\bar{z}_{m,3}}] \nonumber\\
  \bar{z}_{m,1}
  &=
  -\frac{p_m}{1 + p_m v_1} 
  \sum_{\alpha=1}^3 
    (s_{i \alpha} - b_{i \alpha})^2 
  \nonumber\\
  \bar{z}_{m,2}
  &= 
  \left[%
  (1 + p_m v_1) 
  \sum_{\alpha=1}^3 
    \left[(s_{i\alpha} - a_{i\alpha}) + p_m v_1 (b_{i\alpha} - a_{i\alpha}) \right]^2 
  \right]^{\frac{1}{2}}
    \nonumber\\
  \bar{z}_{m,3}
  &= 
  \left[%
  \frac{1}{v_1 (1 + p_m v_1)} 
  \sum_{\alpha=1}^3 
    \left[(s_{i \alpha} - a_{i \alpha})+ p_m v_1 (b_{i \alpha} - a_{i \alpha}) \right]^2 
  \right]^\frac{1}{2} \; .
    \label{eq:riarib-final}
\end{align}

\clearpage
\begin{table}[h]
  \caption{%
    He ($1\ ^1\text{S}$) convergence table.
    \label{tab:hesom}
  }
  \begin{tabular}{@{}l@{\ } lll l @{}}
    \hline\hline\\[-0.40cm]
    \multicolumn{1}{l}{$E_{\text{nr}}$} & 
    \multicolumn{1}{l}{$-\frac{1}{8}\sum_i \left\langle \bos{\nabla}_i^2 \bos{\nabla}_i^2 \right\rangle$}  &
    \multicolumn{1}{l}{$\sum_{i,a} Z_a \left\langle \delta(\br_{ia})\right\rangle$}  &
    \multicolumn{1}{l}{$\sum_{i,j} \left\langle \delta(\br_{ij})\right\rangle$} 
    \\
    \hline\\[-0.35cm]
    \multicolumn{2}{l}{Direct evaluation:} \\ 
    $-$2.903 724 2      & $-$13.50          & 7.229        & 0.106 92        \\    
    $-$2.903 724 36     & $-$13.516         & 7.238        & 0.106 54        \\    
    $-$2.903 724 376 89  & $-$13.519        & 7.240 1      & 0.106 38 \\
    $-$2.903 724 376 968 & $-$13.520 2      & 7.240 6      & 0.106 37 \\
    \hline \\[-0.35cm]
    \multicolumn{3}{l}{Numerical drachmannization:} \\    
    $-$2.903 724 2       & $-$13.522 86   & 7.241 69     & 0.106 342 1      \\
    $-$2.903 724 36      & $-$13.522 32   & 7.241 715    & 0.106 344 9      \\
    $-$2.903 724 376 89  & $-$13.522 071  & 7.241 717 0  & 0.106 345 356  \\
    $-$2.903 724 376 968 & $-$13.522 054  & 7.241 717 15 & 0.106 345 364  \\
    \hline \\[-0.35cm]
    Ref.~\cite{Dr06} \\
    $-$2.903 724 377 034 1195 
    & $-$13.522 016 81
    &     7.241 717 274
    &     0.106 345 371 2 \\
    \hline\hline\\[-0.35cm]
  \end{tabular}  
\end{table}

\begin{table}[h]
  \caption{%
    Li ($2\ ^2\text{S}$) convergence table.
    \label{tab:lisom}
  }
  \begin{tabular}{@{}l@{\ } llll @{}}
    \hline\hline\\[-0.40cm]
    \multicolumn{1}{l}{$E_\text{nr}$} &
    \multicolumn{1}{l}{$-\frac{1}{8}\sum_i \left\langle \bos{\nabla}_i^2 \bos{\nabla}_i^2 \right\rangle$}  &
    \multicolumn{1}{l}{$\sum_{i,a} Z_a \left\langle \delta(\br_{ia})\right\rangle$}  &
    \multicolumn{1}{l}{$\sum_{i,j} \left\langle \delta(\br_{ij})\right\rangle$}
    \\
    \hline\\[-0.35cm]
    \multicolumn{3}{l}{Direct evaluation:} \\          
    $-$7.477 8        & $-$77.5     & 40.9       & 0.557       \\
    $-$7.478 034      & $-$78.2     & 41.3       & 0.552       \\
    $-$7.478 056 9    & $-$78.43    & 41.45      & 0.548 2     \\
    $-$7.478 060 22   & $-$78.46    & 41.46      & 0.545 3     \\
    \hline \\[-0.35cm]
    \multicolumn{3}{l}{Numerical drachmannization:} \\        
    $-$7.477 8        & $-$78.545   & 41.515     & 0.543 8     \\
    $-$7.478 034      & $-$78.565   & 41.526 1   & 0.544 18    \\
    $-$7.478 056 9    & $-$78.561   & 41.527 61  & 0.544 29    \\
    $-$7.478 060 22   & $-$78.557 5 & 41.527 76  & 0.544 322   \\
    \hline \\[-0.35cm]
    Ref.~\cite{YaDr00}$^\ast$ \\
    $-$7.478 060 323 650 3(71) 
        & $-$78.556 143 062(5)
        & 41.527 828 927 (65)
        & 0.544 329 79(31)
        \\
    \hline\hline
  \end{tabular} 
  ~\\
{\footnotesize
  $^\ast$~A more precise non-relativistic energy value, $E_\text{nr}=-7.478\ 060\ 323\ 910 2(2)\ \Eh$, was reported in Ref.~\cite{PuKoPa10}, but we can take the regularized relativistic corrections from Ref.~\cite{YaDr00} for testing our results.
}
\end{table}

\begin{table}[h]
  \caption{%
    Be ($2\ ^1\text{S}$) convergence table.
    \label{tab:besom}
  }
  \begin{tabular}{@{}l@{\ } lll l @{}}
    \hline\hline\\[-0.40cm]
    \multicolumn{1}{l}{$E_\text{nr}$} &
    \multicolumn{1}{l}{$-\frac{1}{8}\sum_i \left\langle \bos{\nabla}_i^2 \bos{\nabla}_i^2 \right\rangle$} &
    \multicolumn{1}{l}{$\sum_{i,a} Z_a \left\langle \delta(\br_{ia})\right\rangle$} &
    \multicolumn{1}{l}{$\sum_{i,j} \left\langle \delta(\br_{ij})\right\rangle$}
    \\
    \hline\\[-0.35cm]
    \multicolumn{3}{l}{Direct evaluation:} \\    
    $-$14.667 26           & $-$269        & 140.9      & 1.624       \\
    $-$14.667 345          & $-$270.21     & 141.18     & 1.613       \\
    $-$14.667 354 1        & $-$270.40     & 141.29     & 1.609 6     \\
    $-$14.667 355 8        & $-$270.41     & 141.30     & 1.607 7     \\    
    $-$14.667 356 03       & $-$270.43     & 141.31     & 1.607 5     \\
    \hline \\[-0.35cm]
    \multicolumn{3}{l}{Numerical drachmannization:} \\    
    $-$14.667 26           & $-$270.72             & 141.471 2  & 1.605 0     \\
    $-$14.667 345          & $-$270.71             & 141.475 2  & 1.605 24    \\
    $-$14.667 354 1        & $-$270.71             & 141.475 68 & 1.605 28    \\
    $-$14.667 355 8        & $-$270.706 7          & 141.475 77 & 1.605 300 4 \\
    $-$14.667 356 03       & $-$270.706 5          & 141.475 80 & 1.605 301 5 \\
    \hline \\[-0.35cm]
    Ref.~\cite{PuKoPa13}$^\ast$ \\
    $-$14.667 356 498
    & $-$270.703 76
    & 141.476 010 4
    & 1.605 305 33 
    \\
    \hline\hline
  \end{tabular} 
  ~\\
  {\footnotesize %
  $^\ast$~A more precise non-relativistic energy was reported in Ref.~\cite{HoAdBu19}, but we cite the Ref.~\cite{PuPaKo14} value because we test our results with respect to the regularized relativistic corrections from that work.
  }

\end{table}

\begin{table}[h]
  \caption{
    $\text{H}_2$ ($\text{X}\ ^1\Sgp, R_\text{p-p}=1.40$~bohr) convergence table.
    \label{tab:h2som}
  }
\scalebox{1.}{%
  \begin{tabular}{@{}l@{\ } l@{\ }l@{\ }l@{\ }l  @{}}
    \hline\hline\\[-0.40cm]
    \multicolumn{1}{l}{$E_\text{nr}$} &
    \multicolumn{1}{l}{$-\frac{1}{8}\sum_i \langle \bos{\nabla}_i^2 \bos{\nabla}_i^2 \rangle$}  &
    \multicolumn{1}{l}{$\sum_{i,a} Z_a \langle \delta(\br_{ia}) \rangle$}  &
    \multicolumn{1}{l}{$\sum_{i,j} \langle \delta(\br_{ij}) \rangle$} &
    \multicolumn{1}{l}{$\langle H_\text{BP}^{(2)} \rangle$} 
    \\
    \hline\\[-0.35cm]
    \multicolumn{3}{l}{Direct evaluation:} \\    
    $-$1.174 475 5     &    $-$1.651 7   & 0.917 5        & 0.016 87      & $-$0.205 \\
    $-$1.174 475 697   &	$-$1.653 3   & 0.918 5        & 0.016 795     & $-$0.205 4 \\
    $-$1.174 475 712 8 &	$-$1.653 4   & 0.918 5        & 0.016 779     & $-$0.205 5 \\
    $-$1.174 475 713 8 &	$-$1.653 5   & 0.918 6        & 0.016 771     & $-$0.205 5 \\
    \hline\\[-0.35cm]
    \multicolumn{3}{l}{Numerical drachmannization:} \\    
    $-$1.174 475 5     &	$-$1.655    & 0.919 323    & 0.016 742   & $-$0.206  \\
    $-$1.174 475 697   & $-$1.654 8  & 0.919 335 5  & 0.016 743 1  & $-$0.205 8 \\
    $-$1.174 475 712 8 & $-$1.654 80 & 0.919 335 7  & 0.016 743 21  & $-$0.205 75 \\
    $-$1.174 475 713 8 & $-$1.654 79 & 0.919 335 8  & 0.016 743 24 & $-$0.205 733 \\    
    \hline\\[-0.35cm]
    Ref.~\cite{PuKoPa17} \\
    $-$1.174 475 714 220 443 4 
    & $-$1.654 744 522 5 
    & 0.919 336 211 2 
    & 0.016 743 278 3 
    & $-$0.205 688 526 
    \\
    \hline\hline
  \end{tabular} 
  }\\
{\footnotesize %
The (non-singular) $\langle H_\text{OO}\rangle$ term is 
 $-$0.047~638,
 $-$0.047~635,
 $-$0.047~634~7,
$-$0.047~634~6, for the increasingly accurate basis sets listed in the table.
The reference value is $-$0.047~634~494(2) \cite{PuKoPa17}.
}
\end{table}

\begin{table}[h]
  \caption{
    $\text{H}_3^{\text{+}}$ ($\text{A}'_1,R_{\text{p-p}}=1.65$~bohr equilateral triangular structure) convergence table.
    \label{tab:h3psom}
  }
\scalebox{1.}{%
  \begin{tabular}{@{}l@{\ } l@{\ }l@{\ }l@{\ }l@{\ }l  @{}}
    \hline\hline\\[-0.40cm]
    \multicolumn{1}{l}{$E_\text{nr}$} &
    \multicolumn{1}{l}{$-\frac{1}{8}\sum_i \langle \bos{\nabla}_i^2 \bos{\nabla}_i^2 \rangle$}  &
    \multicolumn{1}{l}{$\sum_{i,a} Z_a \langle \delta(\br_{ia}) \rangle$}  &
    \multicolumn{1}{l}{$\sum_{i,j} \langle \delta(\br_{ij}) \rangle$} &
    \multicolumn{1}{l}{$\langle H_\text{OO} \rangle$} &
    \multicolumn{1}{l}{$\langle H_\text{BP}^{(2)} \rangle$}
    \\
    \hline\\[-0.35cm]
    \multicolumn{3}{l}{Direct evaluation:} \\    
    $-$1.343 826        & $-$1.920       & 1.082 5     & 0.019 1        & $-$0.057 3       & $-$0.218       \\
    $-$1.343 835 25     &	$-$1.923       & 1.083 9     & 0.018 5        & $-$0.057 22      & $-$0.220       \\
    $-$1.343 835 606    &	$-$1.930 8     & 1.088 1     & 0.018 4        & $-$0.057 218     & $-$0.221 0     \\
    $-$1.343 835 623    &	$-$1.931 9     & 1.088 8     & 0.018 37       & $-$0.057 217 6   & $-$0.221 22    \\
    $-$1.343 835 625 3  &	$-$1.932 0     & 1.088 8     & 0.018 358      & $-$0.057 217 524 & $-$0.221 238   \\
    $-$1.343 835 625 4  &	$-$1.932 0     & 1.088 8     & 0.018 357      & $-$0.057 217 520 & $-$0.221 240 1 \\	
    \multicolumn{3}{l}{Numerical drachmannization:} \\    
    $-$1.343 826        & $-$1.934 5    & 1.089 601    & 0.018 3       & $-$0.057 3       & $-$0.222 8     \\
    $-$1.343 835 25     &	$-$1.933 58   & 1.089 628    & 0.018 333     & $-$0.057 22      & $-$0.221 6     \\
    $-$1.343 835 606    &	$-$1.933 52   & 1.089 652 4  & 0.018 334 4   & $-$0.057 218     & $-$0.221 524   \\
    $-$1.343 835 623    &	$-$1.933 466  & 1.089 654 2  & 0.018 334 63  & $-$0.057 217 6   & $-$0.221 460   \\
    $-$1.343 835 625 3  &	$-$1.933 450  & 1.089 654 22 & 0.018 334 665 & $-$0.057 217 524 & $-$0.221 443   \\
    $-$1.343 835 625 4  & $-$1.933 449 7  & 1.089 654 23 & 0.018 334 666 & $-$0.057 217 520 & $-$0.221 442 3 \\
    \hline\\[-0.35cm]
    Ref.~\cite{JeIrFeMa22} \\
    $-$1.343 835 625 4
    & $-$1.933 424
    & 1.089 655
    & 0.018 335 
    & $-$0.057 218
    & $-$0.221 422
    \\
    \hline\hline
  \end{tabular} 
  }\\
{\footnotesize %
IT: integral transform \cite{JeIrFeMa22,PaCeKo05}. All basis sets were taken from Ref.~\cite{JeIrFeMa22}.
}
\end{table}

\begin{table}[h]
  \caption{%
    H$_3$
    ($1\ ^2\Sigma^+$, 
    $R_{\text{p}_1-\text{p}_2}=1.4015$~bohr, 
    $R_{\text{p}_2-\text{p}_3}=5.8113$~bohr, 
    $\vartheta_{\text{p}_1-\text{p}_2-\text{p}_3}=180^\text{o}$)
    convergence table.
    The (equilibrium) geometry was taken from Ref.~\cite{MiKaPe02}.
    \label{tab:h3som}
  }
\scalebox{1.}{%
  \begin{tabular}{@{}l@{\ } l@{\ }l@{\ }l@{\ }l@{\ }l  @{}}
    \hline\hline\\[-0.40cm]
    \multicolumn{1}{l}{$E_\text{nr}$} &
    \multicolumn{1}{l}{$-\frac{1}{8}\sum_i \langle \bos{\nabla}_i^2 \bos{\nabla}_i^2 \rangle$}  &
    \multicolumn{1}{l}{$\sum_{i,a} Z_a \langle \delta(\br_{ia}) \rangle$}  &
    \multicolumn{1}{l}{$\sum_{i,j} \langle \delta(\br_{ij}) \rangle$} &
    \multicolumn{1}{l}{$\langle H_\text{OO} \rangle$} &
    \multicolumn{1}{l}{$\langle H_\text{BP}^{(2)} \rangle$}
    \\
    \hline\\[-0.35cm]
    \multicolumn{3}{l}{Direct evaluation:} \\   
    $-$1.674 561 1
    & $-$2.272 
    & 1.233
    & 0.016 77
    & $-$0.046 782 
    & $-$0.329 4
    \\
    $-$1.674 562 0 
    & $-$2.274 
    &  1.234 
    &  0.016 78
    & $-$0.046 780 6
    & $-$0.329 35
    \\
    $-$1.674 562 17 
    & $-$2.275 
    &  1.234 5
    &  0.016 78 
    & $-$0.046 780 3
    & $-$0.329 35
    \\
    $-$1.674 562 27 
    & $-$2.275 
    &    1.234 9 
    &    0.016 77 
    & $-$0.046 780 1 
    & $-$0.329 36 
    \\
    \multicolumn{3}{l}{Numerical drachmannization:} \\        
    $-$1.674 561 1
    & $-$2.277 82
    &    1.236 55
    &    0.016 711	
    & $-$0.046 782 
    & $-$0.329 72
    \\
    $-$1.674 562 0
    & $-$2.277 86
    &    1.236 56
    &    0.016 712	
    & $-$0.046 780 6
    & $-$0.329 75
    \\
    $-$1.674 562 17
    & $-$2.277 868	
    &    1.236 565
    &    0.016 712 0
    & $-$0.046 780 3
    & $-$0.329 755
    \\
    $-$1.674 562 27 
    & $-$2.277 870 
    &    1.236 566 
    &    0.016 712 1 
    & $-$0.046 780 1
    & $-$0.329 754 
    \\
    \hline\hline
  \end{tabular} 
}
\end{table}

\begin{table}[h]
  \caption{%
    $\text{HeH}^{+}$ ($1\ ^1\Sigma^+$, $R_{\text{p-}\alpha}=1.46$~bohr) convergence table.
    \label{tab:hehpsom}    
  }
\scalebox{1.}{%
  \begin{tabular}{@{}l@{\ } l@{\ }l@{\ }l@{\ }l@{\ }l  @{}}
    \hline\hline\\[-0.40cm]
    \multicolumn{1}{l}{$E_\text{nr}$} &
    \multicolumn{1}{l}{$-\frac{1}{8}\sum_i \langle \bos{\nabla}_i^2 \bos{\nabla}_i^2 \rangle$}  &
    \multicolumn{1}{l}{$\sum_{i,a} Z_a \langle \delta(\br_{ia}) \rangle$}  &
    \multicolumn{1}{l}{$\sum_{i,j} \langle \delta(\br_{ij}) \rangle$} &
    \multicolumn{1}{l}{$\langle H_\text{OO} \rangle$} &
    \multicolumn{1}{l}{$\langle H_\text{BP}^{(2)} \rangle$}
    \\
    \hline\\[-0.35cm]
    \multicolumn{3}{l}{Direct evaluation:} \\   
    $-$2.978 706 5     & $-$13.403    & 7.206         & 0.101 43      & $-$0.141 244    & $-$1.906 0  \\
    $-$2.978 706 59    & $-$13.406    & 7.208         & 0.101 27      & $-$0.141 242 0  & $-$1.906 4  \\
    $-$2.978 706 597   & $-$13.406    & 7.208         & 0.101 24      & $-$0.141 241 8  & $-$1.906 4  \\
    $-$2.978 706 598 2 & $-$13.406    & 7.208         & 0.101 23      & $-$0.141 241 68 & $-$1.906 5  \\
    $-$2.978 706 598 5 & $-$13.406    & 7.208         & 0.101 22      & $-$0.141 241 61 & $-$1.906 6  \\  
    \multicolumn{3}{l}{Numerical drachmannization:} \\        
    $-$2.978 706 5     & $-$13.420    & 7.216 238     & 0.101 121     & $-$0.141 244    & $-$1.908    	\\
    $-$2.978 706 59    & $-$13.419 6  & 7.216 244     & 0.101 121 6   & $-$0.141 242 0  & $-$1.907 9	\\
    $-$2.978 706 597   & $-$13.419 52 & 7.216 244 5   & 0.101 121 66  & $-$0.141 241 8  & $-$1.907 84	\\
    $-$2.978 706 598 2 & $-$13.419 51 & 7.216 244 67  & 0.101 121 69  & $-$0.141 241 68 & $-$1.907 82  	\\
    $-$2.978 706 598 5 & $-$13.419 497  & 7.216 244 78 & 0.101 121 71 & $-$0.141 241 61 & $-$1.907 804  \\
    \hline\\[-0.35cm]
    Ref.~\cite{JeFeMa21} \\
    $-$2.978 706 598 5
    & $-$13.419 29
    &     7.216 253 
    &     0.101 122 
    &  $-$0.141 241 61 
    &  $-$1.907 581
    \\
    \hline\hline
  \end{tabular} 
}~\\
{\footnotesize %
All basis sets were taken from Ref.~\cite{JeFeMa21}.
}
\end{table}

\begin{table}[h]
  \caption{%
    He$_2^+$ ($1\ ^2\Sigma_\text{u}^+$, $R_{\alpha-\alpha}=2.042$~bohr) convergence table.
    \label{tab:he2psom}
  }
\scalebox{1.}{%
  \begin{tabular}{@{}l@{\ } l@{\ }l@{\ }l@{\ }l@{\ }l  @{}}
    \hline\hline\\[-0.40cm]
    \multicolumn{1}{l}{$E_\text{nr}$} &
    \multicolumn{1}{l}{$-\frac{1}{8}\sum_i \langle \bos{\nabla}_i^2 \bos{\nabla}_i^2 \rangle$}  &
    \multicolumn{1}{l}{$\sum_{i,a} Z_a \langle \delta(\br_{ia}) \rangle$}  &
    \multicolumn{1}{l}{$\sum_{i,j} \langle \delta(\br_{ij}) \rangle$} &
    \multicolumn{1}{l}{$\langle H_\text{OO} \rangle$} &
    \multicolumn{1}{l}{$\langle H_\text{BP}^{(2)} \rangle$}
    \\
    \hline\\[-0.35cm]
    \multicolumn{3}{l}{Direct evaluation:} \\ 
    $-$4.994 637
    &  $-$23.78 
    &   12.48 
    &  0.118 5 
    &  $-$0.086 62  
    &   $-$3.882 
    \\
    $-$4.994 641 
    &  $-$23.79 
    &   12.49 
    &   0.118 3 
    & $-$0.086 60  
    &  $-$3.884 
    \\
    $-$4.994 643 
    & $-$23.79 
    & 12.49 
    & 0.118 3 
    & $-$0.086 59
    & $-$3.884 
    \\
    $-$4.994 644 14 
    & $-$23.83 
    & 12.52 
    & 0.118 0 
    & $-$0.086 577 
    & $-$3.887  \\
    \multicolumn{3}{l}{Numerical drachmannization:} \\        
    $-$4.994 637
    & $-$23.867 8
    &  12.537 3 
    &  0.117 58 
    &  $-$0.086 62 
    &  $-$3.891 4 
    \\
    $-$4.994 641 
    &   $-$23.867 7
    &    12.537 4 
    &    0.117 590 
    &    $-$0.086 60 
    &   $-$3.891 2 
    \\
    $-$4.994 643 
    & $-$23.867 80
    & 12.537 41 
    & 0.117 592 
    & $-$0.086 59 
    & $-$3.891 2  \\
    $-$4.994 644 14 
    &  $-$23.867 58
    &  12.537 54 
    & 0.117 596 4 
    & $-$0.086 577 
    & $-$3.890 79 \\
    \hline\\[-0.35cm]
    Ref.~\cite{FeKoMa20} \\
    $-$4.994 644  00 
    & $-$23.866 14
    &  12.537 51 
    &  0.117 628 
    &  $-$0.086 582 
    &  $-$3.889 30
    \\
    \hline\hline
  \end{tabular} 
  }\\
{\footnotesize %
}
\end{table}

\begin{table}[h]
  \caption{%
    He$_2$ (a(1)$\ ^3\Sigma_\text{u}^+$, $R_{\alpha-\alpha}=2.000$~bohr)
    convergence table.
    \label{tab:he2som}
  }
\scalebox{0.9}{%
  \begin{tabular}{@{}ll@{\ }l@{\ }l@{\ }l@{\ }l  @{}}
    \hline\hline\\[-0.40cm]
    \multicolumn{1}{l}{$E_\text{nr}$} &
    \multicolumn{1}{l}{$-\frac{1}{8}\sum_i \langle \bos{\nabla}_i^2 \bos{\nabla}_i^2 \rangle$}  &
    \multicolumn{1}{l}{$\sum_{i,a} Z_a \langle \delta(\br_{ia}) \rangle$}  &
    \multicolumn{1}{l}{$\sum_{i,j} \langle \delta(\br_{ij}) \rangle$} &
    \multicolumn{1}{l}{$\langle H_\text{OO} \rangle$} &
    \multicolumn{1}{l}{$\langle H_\text{BP}^{(2)} \rangle$}
    \\
    \hline\\[-0.35cm]
    \multicolumn{3}{l}{Direct evaluation:} \\   
    $-$5.150 8
    & $-$23.8
    &  12.42
    &  0.124 
    &  $-$0.073 7 
    &   $-$3.937 
    \\
    $-$5.151 06 
    & $-$24.1 
    & 12.63 
    & 0.120 51
    & $-$0.073 25 
    & $-$3.958 
    \\
    $-$5.151 11 
    & $-$24.13
    & 12.644 
    & 0.120 49
    & $-$0.073 18
    & $-$3.959 
    \\
    $-$5.151 121 7 
    & $-$24.135 
    & 12.645 
    & 0.120 40 
    & $-$0.073 156 
    & $-$3.967 
    \\
    $-$5.151 122 3 
    & $-$24.139 
    &    12.645  
    &     0.120 38 
    &  $-$0.073 154 2
    &  $-$3.972  
    \\
    \multicolumn{3}{l}{Numerical drachmannization:} \\        
    $-$5.150 8
    & $-$24.145 
    &  12.654 
    &  0.119 7
    &  $-$0.073 7
    &   $-$3.965 86 
    \\
    $-$5.151 06 
    & $-$24.156 6 
    &  12.660 89 
    & 0.119 93
    & $-$0.073 25 
    & $-$3.965 44 
    \\
    $-$5.151 11 
    & $-$24.157 29
    & 12.660 97 
    & 0.119 957
    & $-$0.073 18 
    & $-$3.965 80 
    \\
    $-$5.151 121 7
    & $-$24.157 30 
    & 12.660 992
    & 0.119 966 
    & $-$0.073 156 
    & $-$3.965 733 
    \\
    $-$5.151 122 3 
    & $-$24.157 271 
    &    12.660 991 
    &     0.119 967 0 
    &  $-$0.073 154 2 
    &  $-$3.965 700 
    \\   

    \hline\hline
  \end{tabular} 
}
\end{table}


\begin{thebibliography}{39}%
\makeatletter
\providecommand \@ifxundefined [1]{%
 \@ifx{#1\undefined}
}%
\providecommand \@ifnum [1]{%
 \ifnum #1\expandafter \@firstoftwo
 \else \expandafter \@secondoftwo
 \fi
}%
\providecommand \@ifx [1]{%
 \ifx #1\expandafter \@firstoftwo
 \else \expandafter \@secondoftwo
 \fi
}%
\providecommand \natexlab [1]{#1}%
\providecommand \enquote  [1]{``#1''}%
\providecommand \bibnamefont  [1]{#1}%
\providecommand \bibfnamefont [1]{#1}%
\providecommand \citenamefont [1]{#1}%
\providecommand \href@noop [0]{\@secondoftwo}%
\providecommand \href [0]{\begingroup \@sanitize@url \@href}%
\providecommand \@href[1]{\@@startlink{#1}\@@href}%
\providecommand \@@href[1]{\endgroup#1\@@endlink}%
\providecommand \@sanitize@url [0]{\catcode `\\12\catcode `\$12\catcode
  `\&12\catcode `\#12\catcode `\^12\catcode `\_12\catcode `\%12\relax}%
\providecommand \@@startlink[1]{}%
\providecommand \@@endlink[0]{}%
\providecommand \url  [0]{\begingroup\@sanitize@url \@url }%
\providecommand \@url [1]{\endgroup\@href {#1}{\urlprefix }}%
\providecommand \urlprefix  [0]{URL }%
\providecommand \Eprint [0]{\href }%
\providecommand \doibase [0]{https://doi.org/}%
\providecommand \selectlanguage [0]{\@gobble}%
\providecommand \bibinfo  [0]{\@secondoftwo}%
\providecommand \bibfield  [0]{\@secondoftwo}%
\providecommand \translation [1]{[#1]}%
\providecommand \BibitemOpen [0]{}%
\providecommand \bibitemStop [0]{}%
\providecommand \bibitemNoStop [0]{.\EOS\space}%
\providecommand \EOS [0]{\spacefactor3000\relax}%
\providecommand \BibitemShut  [1]{\csname bibitem#1\endcsname}%
\let\auto@bib@innerbib\@empty
\bibitem [{\citenamefont {Bethe}\ and\ \citenamefont
  {Salpeter}(1957)}]{BeSabook57}%
  \BibitemOpen
  \bibfield  {author} {\bibinfo {author} {\bibfnamefont {H.~A.}\ \bibnamefont
  {Bethe}}\ and\ \bibinfo {author} {\bibfnamefont {E.~E.}\ \bibnamefont
  {Salpeter}},\ }\href@noop {} {\emph {\bibinfo {title} {Quantum Mechanics of
  One- and Two-Electron Atoms}}}\ (\bibinfo  {publisher} {Springer},\ \bibinfo
  {address} {Berlin},\ \bibinfo {year} {1957})\BibitemShut {NoStop}%
\bibitem [{\citenamefont {Drachman}(1981)}]{Dr81}%
  \BibitemOpen
  \bibfield  {author} {\bibinfo {author} {\bibfnamefont {R.~J.}\ \bibnamefont
  {Drachman}},\ }\bibfield  {title} {\bibinfo {title} {{A new global operator
  for two-particle delta functions}},\ }\href
  {https://doi.org/10.1088/0022-3700/14/16/003} {\bibfield  {journal} {\bibinfo
   {journal} {J. Phys. B}\ }\textbf {\bibinfo {volume} {14}},\ \bibinfo {pages}
  {2733} (\bibinfo {year} {1981})}\BibitemShut {NoStop}%
\bibitem [{\citenamefont {Drake}(2006)}]{Dr06}%
  \BibitemOpen
  \bibfield  {author} {\bibinfo {author} {\bibfnamefont {G.}~\bibnamefont
  {Drake}},\ }\href {https://doi.org/10.1007/978-0-387-26308-3_11} {\emph
  {\bibinfo {title} {High {P}recision {C}alculations for {H}elium, {In}: Drake
  G. (eds) {S}pringer {H}andbook of {A}tomic, {M}olecular, and {O}ptical
  {P}hysics. {S}pringer {H}andbooks}}}\ (\bibinfo  {publisher} {Springer},\
  \bibinfo {address} {New York, NY},\ \bibinfo {year} {2006})\ pp.\ \bibinfo
  {pages} {199--219}\BibitemShut {NoStop}%
\bibitem [{\citenamefont {Yan}\ and\ \citenamefont {Drake}(2000)}]{YaDr00}%
  \BibitemOpen
  \bibfield  {author} {\bibinfo {author} {\bibfnamefont {Z.-C.}\ \bibnamefont
  {Yan}}\ and\ \bibinfo {author} {\bibfnamefont {G.~W.~F.}\ \bibnamefont
  {Drake}},\ }\bibfield  {title} {\bibinfo {title} {Lithium isotope shifts as a
  measure of nuclear size},\ }\href
  {https://doi.org/10.1103/PhysRevA.61.022504} {\bibfield  {journal} {\bibinfo
  {journal} {Phys. Rev. A}\ }\textbf {\bibinfo {volume} {61}},\ \bibinfo
  {pages} {022504} (\bibinfo {year} {2000})}\BibitemShut {NoStop}%
\bibitem [{\citenamefont {Puchalski}\ \emph {et~al.}(2010)\citenamefont
  {Puchalski}, \citenamefont {Kedziera},\ and\ \citenamefont
  {Pachucki}}]{PuKoPa10}%
  \BibitemOpen
  \bibfield  {author} {\bibinfo {author} {\bibfnamefont {M.}~\bibnamefont
  {Puchalski}}, \bibinfo {author} {\bibfnamefont {D.}~\bibnamefont
  {Kedziera}},\ and\ \bibinfo {author} {\bibfnamefont {K.}~\bibnamefont
  {Pachucki}},\ }\bibfield  {title} {\bibinfo {title} {Ionization potential for
  excited {$S$} states of the lithium atom},\ }\href
  {https://doi.org/10.1103/PhysRevA.82.062509} {\bibfield  {journal} {\bibinfo
  {journal} {Phys. Rev. A}\ }\textbf {\bibinfo {volume} {82}},\ \bibinfo
  {pages} {062509} (\bibinfo {year} {2010})}\BibitemShut {NoStop}%
\bibitem [{\citenamefont {Pachucki}\ \emph {et~al.}(2005)\citenamefont
  {Pachucki}, \citenamefont {Cencek},\ and\ \citenamefont {Komasa}}]{PaCeKo05}%
  \BibitemOpen
  \bibfield  {author} {\bibinfo {author} {\bibfnamefont {K.}~\bibnamefont
  {Pachucki}}, \bibinfo {author} {\bibfnamefont {W.}~\bibnamefont {Cencek}},\
  and\ \bibinfo {author} {\bibfnamefont {J.}~\bibnamefont {Komasa}},\
  }\bibfield  {title} {\bibinfo {title} {On the acceleration of the convergence
  of singular operators in {Gaussian} basis sets},\ }\href
  {https://doi.org/10.1063/1.1888572} {\bibfield  {journal} {\bibinfo
  {journal} {J. Chem. Phys.}\ }\textbf {\bibinfo {volume} {122}},\ \bibinfo
  {pages} {184101} (\bibinfo {year} {2005})}\BibitemShut {NoStop}%
\bibitem [{\citenamefont {Puchalski}\ \emph {et~al.}(2013)\citenamefont
  {Puchalski}, \citenamefont {Komasa},\ and\ \citenamefont
  {Pachucki}}]{PuKoPa13}%
  \BibitemOpen
  \bibfield  {author} {\bibinfo {author} {\bibfnamefont {M.}~\bibnamefont
  {Puchalski}}, \bibinfo {author} {\bibfnamefont {J.}~\bibnamefont {Komasa}},\
  and\ \bibinfo {author} {\bibfnamefont {K.}~\bibnamefont {Pachucki}},\
  }\bibfield  {title} {\bibinfo {title} {Testing quantum electrodynamics in the
  lowest singlet states of the beryllium atom},\ }\href
  {https://doi.org/10.1103/PhysRevA.87.030502} {\bibfield  {journal} {\bibinfo
  {journal} {Phys. Rev. A}\ }\textbf {\bibinfo {volume} {87}},\ \bibinfo
  {pages} {030502} (\bibinfo {year} {2013})}\BibitemShut {NoStop}%
\bibitem [{\citenamefont {Puchalski}\ \emph {et~al.}(2014)\citenamefont
  {Puchalski}, \citenamefont {Pachucki},\ and\ \citenamefont
  {Komasa}}]{PuPaKo14}%
  \BibitemOpen
  \bibfield  {author} {\bibinfo {author} {\bibfnamefont {M.}~\bibnamefont
  {Puchalski}}, \bibinfo {author} {\bibfnamefont {K.}~\bibnamefont
  {Pachucki}},\ and\ \bibinfo {author} {\bibfnamefont {J.}~\bibnamefont
  {Komasa}},\ }\bibfield  {title} {\bibinfo {title} {Isotope shift in a
  beryllium atom},\ }\href {https://doi.org/10.1103/PhysRevA.89.012506}
  {\bibfield  {journal} {\bibinfo  {journal} {Phys. Rev. A}\ }\textbf {\bibinfo
  {volume} {89}},\ \bibinfo {pages} {012506} (\bibinfo {year}
  {2014})}\BibitemShut {NoStop}%
\bibitem [{\citenamefont {Cencek}\ \emph {et~al.}(2005)\citenamefont {Cencek},
  \citenamefont {Komasa}, \citenamefont {Pachucki},\ and\ \citenamefont
  {Szalewicz}}]{CeKoPaSza05}%
  \BibitemOpen
  \bibfield  {author} {\bibinfo {author} {\bibfnamefont {W.}~\bibnamefont
  {Cencek}}, \bibinfo {author} {\bibfnamefont {J.}~\bibnamefont {Komasa}},
  \bibinfo {author} {\bibfnamefont {K.}~\bibnamefont {Pachucki}},\ and\
  \bibinfo {author} {\bibfnamefont {K.}~\bibnamefont {Szalewicz}},\ }\bibfield
  {title} {\bibinfo {title} {Relativistic correction to the helium dimer
  interaction energy},\ }\href
  {https://doi.org/https://doi.org/10.1103/PhysRevLett.95.233004} {\bibfield
  {journal} {\bibinfo  {journal} {Phys. Rev. Lett.}\ }\textbf {\bibinfo
  {volume} {95}},\ \bibinfo {pages} {233004} (\bibinfo {year}
  {2005})}\BibitemShut {NoStop}%
\bibitem [{\citenamefont {Puchalski}\ \emph {et~al.}(2017)\citenamefont
  {Puchalski}, \citenamefont {Komasa},\ and\ \citenamefont
  {Pachucki}}]{PuKoPa17}%
  \BibitemOpen
  \bibfield  {author} {\bibinfo {author} {\bibfnamefont {M.}~\bibnamefont
  {Puchalski}}, \bibinfo {author} {\bibfnamefont {J.}~\bibnamefont {Komasa}},\
  and\ \bibinfo {author} {\bibfnamefont {K.}~\bibnamefont {Pachucki}},\
  }\bibfield  {title} {\bibinfo {title} {Relativistic corrections for the
  ground electronic state of molecular hydrogen},\ }\href
  {https://doi.org/10.1103/PhysRevA.95.052506} {\bibfield  {journal} {\bibinfo
  {journal} {Phys. Rev. A}\ }\textbf {\bibinfo {volume} {95}},\ \bibinfo
  {pages} {052506} (\bibinfo {year} {2017})}\BibitemShut {NoStop}%
\bibitem [{\citenamefont {Jeszenszki}\ \emph {et~al.}(2021)\citenamefont
  {Jeszenszki}, \citenamefont {Ferenc},\ and\ \citenamefont
  {M\'atyus}}]{JeFeMa21}%
  \BibitemOpen
  \bibfield  {author} {\bibinfo {author} {\bibfnamefont {P.}~\bibnamefont
  {Jeszenszki}}, \bibinfo {author} {\bibfnamefont {D.}~\bibnamefont {Ferenc}},\
  and\ \bibinfo {author} {\bibfnamefont {E.}~\bibnamefont {M\'atyus}},\
  }\bibfield  {title} {\bibinfo {title} {All-order explicitly correlated
  relativistic computations for atoms and molecules},\ }\href
  {https://doi.org/10.1063/5.0051237} {\bibfield  {journal} {\bibinfo
  {journal} {J. Chem. Phys.}\ }\textbf {\bibinfo {volume} {154}},\ \bibinfo
  {pages} {224110} (\bibinfo {year} {2021})}\BibitemShut {NoStop}%
\bibitem [{\citenamefont {Jeszenszki}\ \emph
  {et~al.}(2022{\natexlab{a}})\citenamefont {Jeszenszki}, \citenamefont
  {Ireland}, \citenamefont {Ferenc},\ and\ \citenamefont
  {Mátyus}}]{JeIrFeMa22}%
  \BibitemOpen
  \bibfield  {author} {\bibinfo {author} {\bibfnamefont {P.}~\bibnamefont
  {Jeszenszki}}, \bibinfo {author} {\bibfnamefont {R.~T.}\ \bibnamefont
  {Ireland}}, \bibinfo {author} {\bibfnamefont {D.}~\bibnamefont {Ferenc}},\
  and\ \bibinfo {author} {\bibfnamefont {E.}~\bibnamefont {Mátyus}},\
  }\bibfield  {title} {\bibinfo {title} {On the inclusion of cusp effects in
  expectation values with explicitly correlated {G}aussians},\ }\href
  {https://doi.org/https://doi.org/10.1002/qua.26819} {\bibfield  {journal}
  {\bibinfo  {journal} {Int. J. Quant. Chem.}\ }\textbf {\bibinfo {volume}
  {122}},\ \bibinfo {pages} {e26819} (\bibinfo {year}
  {2022}{\natexlab{a}})}\BibitemShut {NoStop}%
\bibitem [{\citenamefont {Ferenc}\ \emph {et~al.}(2020)\citenamefont {Ferenc},
  \citenamefont {Korobov},\ and\ \citenamefont {M\'atyus}}]{FeKoMa20}%
  \BibitemOpen
  \bibfield  {author} {\bibinfo {author} {\bibfnamefont {D.}~\bibnamefont
  {Ferenc}}, \bibinfo {author} {\bibfnamefont {V.~I.}\ \bibnamefont
  {Korobov}},\ and\ \bibinfo {author} {\bibfnamefont {E.}~\bibnamefont
  {M\'atyus}},\ }\bibfield  {title} {\bibinfo {title} {Nonadiabatic,
  relativistic, and leading-order {QED} corrections for rovibrational intervals
  of ${^{4}\mathrm{He}}_{2}^{+}$ (${X}\text{
  }{^{2}\mathrm{\ensuremath{\Sigma}}}_{\mathrm{u}}^{+}$)},\ }\href
  {https://doi.org/10.1103/PhysRevLett.125.213001} {\bibfield  {journal}
  {\bibinfo  {journal} {Phys. Rev. Lett.}\ }\textbf {\bibinfo {volume} {125}},\
  \bibinfo {pages} {213001} (\bibinfo {year} {2020})}\BibitemShut {NoStop}%
\bibitem [{\citenamefont {Saly}\ \emph {et~al.}(2023)\citenamefont {Saly},
  \citenamefont {Ferenc},\ and\ \citenamefont {Mátyus.}}]{SaFeMa22}%
  \BibitemOpen
  \bibfield  {author} {\bibinfo {author} {\bibfnamefont {E.}~\bibnamefont
  {Saly}}, \bibinfo {author} {\bibfnamefont {D.}~\bibnamefont {Ferenc}},\ and\
  \bibinfo {author} {\bibfnamefont {E.}~\bibnamefont {Mátyus.}},\ }\bibfield
  {title} {\bibinfo {title} {Pre-{B}orn--{O}ppenheimer energies, leading-order
  relativistic and {QED} corrections for electronically excited states of
  molecular hydrogen},\ }\href {https://doi.org/10.1080/00268976.2022.2163714}
  {\bibfield  {journal} {\bibinfo  {journal} {Mol. Phys.}\ }\textbf {\bibinfo
  {volume} {121}},\ \bibinfo {pages} {e2163714} (\bibinfo {year}
  {2023})}\BibitemShut {NoStop}%
\bibitem [{\citenamefont {Jeszenszki}\ \emph
  {et~al.}(2022{\natexlab{b}})\citenamefont {Jeszenszki}, \citenamefont
  {Ferenc},\ and\ \citenamefont {M\'atyus}}]{JeFeMa22}%
  \BibitemOpen
  \bibfield  {author} {\bibinfo {author} {\bibfnamefont {P.}~\bibnamefont
  {Jeszenszki}}, \bibinfo {author} {\bibfnamefont {D.}~\bibnamefont {Ferenc}},\
  and\ \bibinfo {author} {\bibfnamefont {E.}~\bibnamefont {M\'atyus}},\
  }\bibfield  {title} {\bibinfo {title} {Variational {D}irac--{C}oulomb
  explicitly correlated computations for molecules},\ }\href
  {https://doi.org/10.1063/5.0075096} {\bibfield  {journal} {\bibinfo
  {journal} {J. Chem. Phys.}\ }\textbf {\bibinfo {volume} {156}},\ \bibinfo
  {pages} {084111} (\bibinfo {year} {2022}{\natexlab{b}})}\BibitemShut
  {NoStop}%
\bibitem [{\citenamefont {Ferenc}\ \emph
  {et~al.}(2022{\natexlab{a}})\citenamefont {Ferenc}, \citenamefont
  {Jeszenszki},\ and\ \citenamefont {M\'atyus}}]{FeJeMa22}%
  \BibitemOpen
  \bibfield  {author} {\bibinfo {author} {\bibfnamefont {D.}~\bibnamefont
  {Ferenc}}, \bibinfo {author} {\bibfnamefont {P.}~\bibnamefont {Jeszenszki}},\
  and\ \bibinfo {author} {\bibfnamefont {E.}~\bibnamefont {M\'atyus}},\
  }\bibfield  {title} {\bibinfo {title} {On the {B}reit interaction in an
  explicitly correlated variational {D}irac--{C}oulomb framework},\ }\href
  {https://doi.org/10.1063/5.0075097} {\bibfield  {journal} {\bibinfo
  {journal} {J. Chem. Phys.}\ }\textbf {\bibinfo {volume} {156}},\ \bibinfo
  {pages} {084110} (\bibinfo {year} {2022}{\natexlab{a}})}\BibitemShut
  {NoStop}%
\bibitem [{\citenamefont {Ferenc}\ \emph
  {et~al.}(2022{\natexlab{b}})\citenamefont {Ferenc}, \citenamefont
  {Jeszenszki},\ and\ \citenamefont {Matyus}}]{FeJeMa22b}%
  \BibitemOpen
  \bibfield  {author} {\bibinfo {author} {\bibfnamefont {D.}~\bibnamefont
  {Ferenc}}, \bibinfo {author} {\bibfnamefont {P.}~\bibnamefont {Jeszenszki}},\
  and\ \bibinfo {author} {\bibfnamefont {E.}~\bibnamefont {Matyus}},\
  }\bibfield  {title} {\bibinfo {title} {Variational versus perturbative
  relativistic energies for small and light atomic and molecular systems},\
  }\href {https://doi.org/10.1063/5.0105355} {\bibfield  {journal} {\bibinfo
  {journal} {J. Chem. Phys.}\ }\textbf {\bibinfo {volume} {157}},\ \bibinfo
  {pages} {094113} (\bibinfo {year} {2022}{\natexlab{b}})}\BibitemShut
  {NoStop}%
\bibitem [{\citenamefont {{J}eszenszki}\ and\ \citenamefont
  {{M}átyus}(2023)}]{JeMa23}%
  \BibitemOpen
  \bibfield  {author} {\bibinfo {author} {\bibfnamefont {P.}~\bibnamefont
  {{J}eszenszki}}\ and\ \bibinfo {author} {\bibfnamefont {E.}~\bibnamefont
  {{M}átyus}},\ }\bibfield  {title} {\bibinfo {title} {{R}elativistic
  two-electron atomic and molecular energies using {$LS$} coupling and double
  groups: role of the triplet contributions to singlet states},\ }\href
  {https://doi.org/10.1063/5.0136360} {\bibfield  {journal} {\bibinfo
  {journal} {J. Chem. Phys.}\ }\textbf {\bibinfo {volume} {158}},\ \bibinfo
  {pages} {054104} (\bibinfo {year} {2023})}\BibitemShut {NoStop}%
\bibitem [{\citenamefont {Ferenc}\ and\ \citenamefont
  {M\'atyus}(2023)}]{FeMa23}%
  \BibitemOpen
  \bibfield  {author} {\bibinfo {author} {\bibfnamefont {D.}~\bibnamefont
  {Ferenc}}\ and\ \bibinfo {author} {\bibfnamefont {E.}~\bibnamefont
  {M\'atyus}},\ }\bibfield  {title} {\bibinfo {title}
  {{P}re--{B}orn-{O}ppenheimer {D}irac-{C}oulomb-{B}reit computations for
  two-body systems},\ }\href {https://doi.org/10.1103/PhysRevA.107.052803}
  {\bibfield  {journal} {\bibinfo  {journal} {Phys. Rev. A}\ }\textbf {\bibinfo
  {volume} {107}},\ \bibinfo {pages} {052803} (\bibinfo {year}
  {2023})}\BibitemShut {NoStop}%
\bibitem [{\citenamefont {M\'atyus}\ \emph {et~al.}(2023)\citenamefont
  {M\'atyus}, \citenamefont {Ferenc}, \citenamefont {Jeszenszki},\ and\
  \citenamefont {Marg\'ocsy}}]{MaFeJeMa23}%
  \BibitemOpen
  \bibfield  {author} {\bibinfo {author} {\bibfnamefont {E.}~\bibnamefont
  {M\'atyus}}, \bibinfo {author} {\bibfnamefont {D.}~\bibnamefont {Ferenc}},
  \bibinfo {author} {\bibfnamefont {P.}~\bibnamefont {Jeszenszki}},\ and\
  \bibinfo {author} {\bibfnamefont {A.}~\bibnamefont {Marg\'ocsy}},\ }\bibfield
   {title} {\bibinfo {title} {The {B}ethe–{S}alpeter {QED} wave equation for
  bound-state computations of atoms and molecules},\ }\href
  {https://doi.org/https://doi.org/10.1021/acsphyschemau.2c00062} {\bibfield
  {journal} {\bibinfo  {journal} {ACS Phys. Chem Au}\ }\textbf {\bibinfo
  {volume} {3}},\ \bibinfo {pages} {222} (\bibinfo {year} {2023})}\BibitemShut
  {NoStop}%
\bibitem [{\citenamefont {Margócsy}\ and\ \citenamefont
  {Mátyus}(2024)}]{MaMa24}%
  \BibitemOpen
  \bibfield  {author} {\bibinfo {author} {\bibfnamefont {A.}~\bibnamefont
  {Margócsy}}\ and\ \bibinfo {author} {\bibfnamefont {E.}~\bibnamefont
  {Mátyus}},\ }\bibfield  {title} {\bibinfo {title} {{QED} corrections to the
  correlated relativistic energy: one-photon processes},\ }\href
  {https://doi.org/10.48550/arXiv.2312.13887} {\  (\bibinfo {year}
  {2024})}\BibitemShut {NoStop}%
\bibitem [{\citenamefont {Nonn}\ \emph {et~al.}(2024)\citenamefont {Nonn},
  \citenamefont {Margócsy},\ and\ \citenamefont {Mátyus}}]{NoMaMa24}%
  \BibitemOpen
  \bibfield  {author} {\bibinfo {author} {\bibfnamefont {A.}~\bibnamefont
  {Nonn}}, \bibinfo {author} {\bibfnamefont {A.}~\bibnamefont {Margócsy}},\
  and\ \bibinfo {author} {\bibfnamefont {E.}~\bibnamefont {Mátyus}},\
  }\href@noop {} {\bibinfo {title} {Bound-state relativistic quantum
  electrodynamics: a perspective for precision physics with atoms and
  molecules, (under review)}} (\bibinfo {year} {2024})\BibitemShut {NoStop}%
\bibitem [{\citenamefont {Rychlewski}(2003)}]{Ry03}%
  \BibitemOpen
  \bibinfo {editor} {\bibfnamefont {J.}~\bibnamefont {Rychlewski}},\ ed.,\
  \href@noop {} {\emph {\bibinfo {title} {Explicitly Correlated Wave Functions
  in Chemistry and Physics}}}\ (\bibinfo  {publisher} {Kluwer Academic
  Publishers},\ \bibinfo {address} {Dodrecht},\ \bibinfo {year}
  {2003})\BibitemShut {NoStop}%
\bibitem [{\citenamefont {Martinazzo}\ and\ \citenamefont
  {Pollak}(2020)}]{MaPo20}%
  \BibitemOpen
  \bibfield  {author} {\bibinfo {author} {\bibfnamefont {R.}~\bibnamefont
  {Martinazzo}}\ and\ \bibinfo {author} {\bibfnamefont {E.}~\bibnamefont
  {Pollak}},\ }\bibfield  {title} {\bibinfo {title} {{Lower bounds to
  eigenvalues of the Schr{\"o}dinger equation by solution of a 90-y
  challenge}},\ }\href {https://doi.org/10.1073/pnas.2007093117} {\bibfield
  {journal} {\bibinfo  {journal} {Proc. Natl. Acad. Sci. U.S.A.}\ }\textbf
  {\bibinfo {volume} {117}},\ \bibinfo {pages} {16181} (\bibinfo {year}
  {2020})}\BibitemShut {NoStop}%
\bibitem [{\citenamefont {Pollak}\ and\ \citenamefont
  {Martinazzo}(2021)}]{PoMa21}%
  \BibitemOpen
  \bibfield  {author} {\bibinfo {author} {\bibfnamefont {E.}~\bibnamefont
  {Pollak}}\ and\ \bibinfo {author} {\bibfnamefont {R.}~\bibnamefont
  {Martinazzo}},\ }\bibfield  {title} {\bibinfo {title} {{Lower Bounds for
  {C}oulombic Systems}},\ }\href {https://doi.org/10.1021/acs.jctc.0c01301}
  {\bibfield  {journal} {\bibinfo  {journal} {J. Chem. Theory Comput.}\
  }\textbf {\bibinfo {volume} {17}},\ \bibinfo {pages} {1535} (\bibinfo {year}
  {2021})}\BibitemShut {NoStop}%
\bibitem [{\citenamefont {Ireland}\ \emph {et~al.}(2021)\citenamefont
  {Ireland}, \citenamefont {Jeszenszki}, \citenamefont {M\'atyus},
  \citenamefont {Martinazzo}, \citenamefont {Ronto},\ and\ \citenamefont
  {Pollak}}]{IrJeMaMaRoPo21}%
  \BibitemOpen
  \bibfield  {author} {\bibinfo {author} {\bibfnamefont {R.}~\bibnamefont
  {Ireland}}, \bibinfo {author} {\bibfnamefont {P.}~\bibnamefont {Jeszenszki}},
  \bibinfo {author} {\bibfnamefont {E.}~\bibnamefont {M\'atyus}}, \bibinfo
  {author} {\bibfnamefont {R.}~\bibnamefont {Martinazzo}}, \bibinfo {author}
  {\bibfnamefont {M.}~\bibnamefont {Ronto}},\ and\ \bibinfo {author}
  {\bibfnamefont {E.}~\bibnamefont {Pollak}},\ }\bibfield  {title} {\bibinfo
  {title} {Lower bounds for atomic energies},\ }\href
  {https://doi.org/10.1021/acsphyschemau.1c00018} {\bibfield  {journal}
  {\bibinfo  {journal} {ACS Phys. Chem. Au}\ }\textbf {\bibinfo {volume} {2}},\
  \bibinfo {pages} {23–37} (\bibinfo {year} {2021})}\BibitemShut {NoStop}%
\bibitem [{\citenamefont {Ronto}\ \emph {et~al.}(2023)\citenamefont {Ronto},
  \citenamefont {Jeszenszki}, \citenamefont {M\'atyus},\ and\ \citenamefont
  {Pollak}}]{RoJeMaPo23}%
  \BibitemOpen
  \bibfield  {author} {\bibinfo {author} {\bibfnamefont {M.}~\bibnamefont
  {Ronto}}, \bibinfo {author} {\bibfnamefont {P.}~\bibnamefont {Jeszenszki}},
  \bibinfo {author} {\bibfnamefont {E.}~\bibnamefont {M\'atyus}},\ and\
  \bibinfo {author} {\bibfnamefont {E.}~\bibnamefont {Pollak}},\ }\bibfield
  {title} {\bibinfo {title} {Lower bounds on par with upper bounds for
  few-electron atomic energies},\ }\href
  {https://doi.org/10.1103/PhysRevA.107.012204} {\bibfield  {journal} {\bibinfo
   {journal} {Phys. Rev. A}\ }\textbf {\bibinfo {volume} {107}},\ \bibinfo
  {pages} {012204} (\bibinfo {year} {2023})}\BibitemShut {NoStop}%
\bibitem [{\citenamefont {Adamowicz}\ \emph {et~al.}(2022)\citenamefont
  {Adamowicz}, \citenamefont {Kvaal}, \citenamefont {Lasser},\ and\
  \citenamefont {Pedersen}}]{AdKvLaPe22}%
  \BibitemOpen
  \bibfield  {author} {\bibinfo {author} {\bibfnamefont {L.}~\bibnamefont
  {Adamowicz}}, \bibinfo {author} {\bibfnamefont {S.}~\bibnamefont {Kvaal}},
  \bibinfo {author} {\bibfnamefont {C.}~\bibnamefont {Lasser}},\ and\ \bibinfo
  {author} {\bibfnamefont {T.~B.}\ \bibnamefont {Pedersen}},\ }\bibfield
  {title} {\bibinfo {title} {{Laser-induced dynamic alignment of the {HD}
  molecule without the {B}orn-{O}ppenheimer approximation}},\ }\href
  {https://doi.org/10.1063/5.0101352} {\bibfield  {journal} {\bibinfo
  {journal} {J. Chem. Phys.}\ }\textbf {\bibinfo {volume} {157}},\ \bibinfo
  {pages} {144302} (\bibinfo {year} {2022})}\BibitemShut {NoStop}%
\bibitem [{\citenamefont {Beutel}\ \emph {et~al.}(2021)\citenamefont {Beutel},
  \citenamefont {Ahrens}, \citenamefont {Huang}, \citenamefont {Suzuki},\ and\
  \citenamefont {Varga}}]{BeAhHuSuVa21}%
  \BibitemOpen
  \bibfield  {author} {\bibinfo {author} {\bibfnamefont {M.}~\bibnamefont
  {Beutel}}, \bibinfo {author} {\bibfnamefont {A.}~\bibnamefont {Ahrens}},
  \bibinfo {author} {\bibfnamefont {C.}~\bibnamefont {Huang}}, \bibinfo
  {author} {\bibfnamefont {Y.}~\bibnamefont {Suzuki}},\ and\ \bibinfo {author}
  {\bibfnamefont {K.}~\bibnamefont {Varga}},\ }\bibfield  {title} {\bibinfo
  {title} {{Deformed explicitly correlated {G}aussians}},\ }\href
  {https://doi.org/10.1063/5.0066427} {\bibfield  {journal} {\bibinfo
  {journal} {J. Chem. Phys.}\ }\textbf {\bibinfo {volume} {155}},\ \bibinfo
  {pages} {214103} (\bibinfo {year} {2021})}\BibitemShut {NoStop}%
\bibitem [{\citenamefont {Beylkin}\ and\ \citenamefont {Lucas}(2005)}]{BeMo05}%
  \BibitemOpen
  \bibfield  {author} {\bibinfo {author} {\bibfnamefont {G.}~\bibnamefont
  {Beylkin}}\ and\ \bibinfo {author} {\bibfnamefont {M.}~\bibnamefont
  {Lucas}},\ }\bibfield  {title} {\bibinfo {title} {On approximation of
  functions by exponential sums},\ }\href
  {https://doi.org/10.1016/j.acha.2005.01.003} {\bibfield  {journal} {\bibinfo
  {journal} {Appl. Comput. Harmon. Anal.}\ }\textbf {\bibinfo {volume} {19}},\
  \bibinfo {pages} {17} (\bibinfo {year} {2005})}\BibitemShut {NoStop}%
\bibitem [{\citenamefont {Paunz}(1979)}]{PaBook79}%
  \BibitemOpen
  \bibfield  {author} {\bibinfo {author} {\bibfnamefont {R.}~\bibnamefont
  {Paunz}},\ }\href@noop {} {\emph {\bibinfo {title} {Spin Eigenfunctions}}}\
  (\bibinfo  {publisher} {Plenum Press},\ \bibinfo {address} {New York},\
  \bibinfo {year} {1979})\BibitemShut {NoStop}%
\bibitem [{\citenamefont {M\'atyus}\ and\ \citenamefont
  {Reiher}(2012)}]{MaRe12}%
  \BibitemOpen
  \bibfield  {author} {\bibinfo {author} {\bibfnamefont {E.}~\bibnamefont
  {M\'atyus}}\ and\ \bibinfo {author} {\bibfnamefont {M.}~\bibnamefont
  {Reiher}},\ }\bibfield  {title} {\bibinfo {title} {Molecular structure
  calculations: {A} unified quantum mechanical description of electrons and
  nuclei using explicitly correlated {G}aussian functions and the global vector
  representation},\ }\href {https://doi.org/10.1063/1.4731696} {\bibfield
  {journal} {\bibinfo  {journal} {J. Chem. Phys.}\ }\textbf {\bibinfo {volume}
  {137}},\ \bibinfo {pages} {024104} (\bibinfo {year} {2012})}\BibitemShut
  {NoStop}%
\bibitem [{\citenamefont {Harrison}\ \emph {et~al.}(2003)\citenamefont
  {Harrison}, \citenamefont {Fann}, \citenamefont {Yanai},\ and\ \citenamefont
  {Beylkin}}]{HaFaYaBe03}%
  \BibitemOpen
  \bibfield  {author} {\bibinfo {author} {\bibfnamefont {R.~J.}\ \bibnamefont
  {Harrison}}, \bibinfo {author} {\bibfnamefont {G.~I.}\ \bibnamefont {Fann}},
  \bibinfo {author} {\bibfnamefont {T.}~\bibnamefont {Yanai}},\ and\ \bibinfo
  {author} {\bibfnamefont {G.}~\bibnamefont {Beylkin}},\ }\bibfield  {title}
  {\bibinfo {title} {Multiresolution {Q}uantum {C}hemistry in {M}ultiwavelet
  {B}ases},\ }in\ \href@noop {} {\emph {\bibinfo {booktitle} {Lecture Notes in
  Computer Science}}},\ Vol.\ \bibinfo {volume} {2660}\ (\bibinfo  {publisher}
  {Springer Berlin Heidelberg},\ \bibinfo {address} {Berlin, Heidelberg},\
  \bibinfo {year} {2003})\ pp.\ \bibinfo {pages} {103--110}\BibitemShut
  {NoStop}%
\bibitem [{\citenamefont {Howells}\ and\ \citenamefont
  {Kennedy}(1990)}]{HoKe90}%
  \BibitemOpen
  \bibfield  {author} {\bibinfo {author} {\bibfnamefont {M.~H.}\ \bibnamefont
  {Howells}}\ and\ \bibinfo {author} {\bibfnamefont {R.~A.}\ \bibnamefont
  {Kennedy}},\ }\bibfield  {title} {\bibinfo {title} {Relativistic corrections
  for the ground and first excited states of {H}$_2^+$, {HD}$^+$ and
  {D}$_2^+$},\ }\href {https://doi.org/10.1039/FT9908603495} {\bibfield
  {journal} {\bibinfo  {journal} {J. Chem. Soc., Faraday Trans.}\ }\textbf
  {\bibinfo {volume} {86}},\ \bibinfo {pages} {3495} (\bibinfo {year}
  {1990})}\BibitemShut {NoStop}%
\bibitem [{\citenamefont {Horny\'ak}\ \emph {et~al.}(2019)\citenamefont
  {Horny\'ak}, \citenamefont {Adamowicz},\ and\ \citenamefont
  {Bubin}}]{HoAdBu19}%
  \BibitemOpen
  \bibfield  {author} {\bibinfo {author} {\bibfnamefont {I.}~\bibnamefont
  {Horny\'ak}}, \bibinfo {author} {\bibfnamefont {L.}~\bibnamefont
  {Adamowicz}},\ and\ \bibinfo {author} {\bibfnamefont {S.}~\bibnamefont
  {Bubin}},\ }\bibfield  {title} {\bibinfo {title} {Ground and excited
  {$^{1}S$} states of the beryllium atom},\ }\href
  {https://doi.org/10.1103/PhysRevA.100.032504} {\bibfield  {journal} {\bibinfo
   {journal} {Phys. Rev. A}\ }\textbf {\bibinfo {volume} {100}},\ \bibinfo
  {pages} {032504} (\bibinfo {year} {2019})}\BibitemShut {NoStop}%
\bibitem [{\citenamefont {Beyer}\ and\ \citenamefont {Merkt}(2016)}]{BeMe16}%
  \BibitemOpen
  \bibfield  {author} {\bibinfo {author} {\bibfnamefont {M.}~\bibnamefont
  {Beyer}}\ and\ \bibinfo {author} {\bibfnamefont {F.}~\bibnamefont {Merkt}},\
  }\bibfield  {title} {\bibinfo {title} {Structure and dynamics of {H}$_2^+$
  near the dissociation threshold: {A} combined experimental and computational
  investigation},\ }\href
  {https://doi.org/https://doi.org/10.1016/j.jms.2016.08.001} {\bibfield
  {journal} {\bibinfo  {journal} {J. Mol. Spectrosc.}\ }\textbf {\bibinfo
  {volume} {330}},\ \bibinfo {pages} {147} (\bibinfo {year}
  {2016})}\BibitemShut {NoStop}%
\bibitem [{\citenamefont {Suzuki}\ and\ \citenamefont
  {Varga}(1998)}]{SuVaBook98}%
  \BibitemOpen
  \bibfield  {author} {\bibinfo {author} {\bibfnamefont {Y.}~\bibnamefont
  {Suzuki}}\ and\ \bibinfo {author} {\bibfnamefont {K.}~\bibnamefont {Varga}},\
  }\href {https://doi.org/10.1007/3-540-49541-X} {\emph {\bibinfo {title}
  {{S}tochastic {V}ariational {A}pproach to {Q}uantum{-}{M}echanical
  {F}ew{-}{B}ody {P}roblems}}}\ (\bibinfo  {publisher} {Springer-Verlag},\
  \bibinfo {address} {Berlin, Heidelberg},\ \bibinfo {year} {1998})\BibitemShut
  {NoStop}%
\bibitem [{\citenamefont {Stanke}\ \emph {et~al.}(2016)\citenamefont {Stanke},
  \citenamefont {Palikot},\ and\ \citenamefont {Adamowicz}}]{StPaAd16}%
  \BibitemOpen
  \bibfield  {author} {\bibinfo {author} {\bibfnamefont {M.}~\bibnamefont
  {Stanke}}, \bibinfo {author} {\bibfnamefont {E.}~\bibnamefont {Palikot}},\
  and\ \bibinfo {author} {\bibfnamefont {L.}~\bibnamefont {Adamowicz}},\
  }\bibfield  {title} {\bibinfo {title} {Algorithms for calculating
  mass-velocity and {Darwin} relativistic corrections with $n$-electron
  explicitly correlated {Gaussians} with shifted centers},\ }\href
  {https://doi.org/10.1063/1.4947553} {\bibfield  {journal} {\bibinfo
  {journal} {J. Chem. Phys.}\ }\textbf {\bibinfo {volume} {144}},\ \bibinfo
  {pages} {174101} (\bibinfo {year} {2016})}\BibitemShut {NoStop}%
\bibitem [{\citenamefont {Mielke}\ \emph {et~al.}(2002)\citenamefont {Mielke},
  \citenamefont {Garrett},\ and\ \citenamefont {Peterson}}]{MiKaPe02}%
  \BibitemOpen
  \bibfield  {author} {\bibinfo {author} {\bibfnamefont {S.~L.}\ \bibnamefont
  {Mielke}}, \bibinfo {author} {\bibfnamefont {B.~C.}\ \bibnamefont
  {Garrett}},\ and\ \bibinfo {author} {\bibfnamefont {K.~A.}\ \bibnamefont
  {Peterson}},\ }\bibfield  {title} {\bibinfo {title} {{A hierarchical family
  of global analytic {B}orn–{O}ppenheimer potential energy surfaces for the
  {H+H}$_2$ reaction ranging in quality from double-zeta to the complete basis
  set limit}},\ }\href {https://doi.org/10.1063/1.1432319} {\bibfield
  {journal} {\bibinfo  {journal} {J. Chem. Phys.}\ }\textbf {\bibinfo {volume}
  {116}},\ \bibinfo {pages} {4142} (\bibinfo {year} {2002})}\BibitemShut
  {NoStop}%
\end{thebibliography}
\end{document}